\documentclass[10pt,journal]{IEEEtran}

\ifCLASSOPTIONcompsoc
  \usepackage[compress]{cite}
\else
  \usepackage{cite}
\fi

\usepackage{ragged2e}
\usepackage{underscore}
\usepackage{pifont}
\usepackage{adjustbox}
\usepackage{hyperref}
\usepackage{amssymb}
\usepackage{graphicx}
\usepackage{multirow}
\usepackage{bbding}
\usepackage{float}
\usepackage{empheq}
\usepackage{amsmath}
\usepackage{cancel}
\usepackage{url}
\usepackage{graphicx}
\usepackage{color}
\usepackage[vlined,ruled]{algorithm2e}
\SetAlgoNlRelativeSize{0} 

\usepackage{setspace}
\usepackage{stfloats}
\usepackage{multirow}
\usepackage{makecell}
\usepackage{multicol}
\usepackage[noend]{algpseudocode}
\SetAlgoNoLine

\algdef{SE}[PROCEDURE]{Procedure}{EndProcedure}[2]{\textbf{procedure} #1(#2)}{\textbf{end procedure}}

\usepackage{bm}
\usepackage{float}
\usepackage[table,xcdraw]{xcolor}
\usepackage[normalem]{ulem}

\usepackage{amsmath}

\everymath{\displaystyle}
\everydisplay{\displaystyle}

\let\oldsum\sum
\renewcommand{\sum}{\displaystyle\oldsum}

\useunder{\uline}{\ul}{}

\usepackage{float}  
\usepackage{lipsum}


\ifCLASSINFOpdf
\else
\fi
\begin{document}
%

\title{
Beyond Model Scale Limits: End-Edge-Cloud Federated Learning with Self-Rectified Knowledge Agglomeration}
%
%
%
%

\author{Zhiyuan~Wu,~\IEEEmembership{Member,~IEEE,}
        Sheng~Sun,
        Yuwei~Wang,~\IEEEmembership{Member,~IEEE,}
        Min~Liu,~\IEEEmembership{Senior~Member,~IEEE,}\\
        Ke~Xu,~\IEEEmembership{Fellow,~IEEE,}
        Quyang~Pan,
        Bo~Gao,~\IEEEmembership{Member,~IEEE},   and~Tian~Wen
\IEEEcompsocitemizethanks{
\IEEEcompsocthanksitem Zhiyuan Wu, Quyang Pan and Tian Wen are with the State Key Lab of Processers, Institute of Computing Technology, Chinese Academy of Sciences, Beijing, China, and also with the University of Chinese Academy of Sciences, Beijing, China.
E-mails: wuzhiyuan22s@ict.ac.cn, panquyang23s@ict.ac.cn, wentian24s@ict.ac.cn. 
\IEEEcompsocthanksitem Sheng Sun and Yuwei Wang are with the State Key Lab of Processers, Institute of Computing Technology, Chinese Academy of Sciences, Beijing, China.
E-mails: \{sunsheng, ywwang\}@ict.ac.cn.
\IEEEcompsocthanksitem Min Liu is with the State Key Lab of Processers, Institute of Computing Technology, Chinese Academy of Sciences, Beijing, China, and also with the Zhongguancun Laboratory, Beijing, China.
E-mail: liumin@ict.ac.cn
\IEEEcompsocthanksitem Ke Xu is with the Department of Computer Science and Technology, Tsinghua University, Beijing, China, and also with the Zhongguancun Laboratory, Beijing, China.
E-mail: xuke@tsinghua.edu.cn.
\IEEEcompsocthanksitem Bo Gao is with the School of Computer Science and Technology, and the Engineering Research Center of Network Management Technology for High-Speed Railway of Ministry of Education, Beijing Jiaotong University, Beijing, China.
E-mail: bogao@bjtu.edu.cn.
\IEEEcompsocthanksitem Corresponding author: Yuwei Wang.
}
\thanks{A preliminary version of this paper titled "Agglomerative Federated Learning: Empowering Larger Model Training via End-Edge-Cloud Collaboration" was accepted by IEEE International Conference on Computer Communications (INFOCOM), 2024. This work was supported by the National Key Research and Development Program of China (No. 2023YFB2703701), the National Natural Science Foundation of China (No. 62472410).}
}
%
%

\markboth{Under Review}%
{Shell \MakeLowercase{\textit{et al.}}: Bare Demo of IEEEtran.cls for Computer Society Journals}
%




\IEEEtitleabstractindextext{%
\begin{abstract}
\justifying
The rise of End-Edge-Cloud Collaboration (EECC) offers a promising paradigm for Artificial Intelligence (AI) model training across end devices, edge servers, and cloud data centers, providing enhanced reliability and reduced latency. Hierarchical Federated Learning (HFL) can benefit from this paradigm by enabling multi-tier model aggregation across distributed computing nodes. 
\textcolor{black}{However, the potential of HFL is significantly constrained by the inherent heterogeneity and dynamic characteristics of EECC environments. Specifically, the uniform model structure bounded by the least powerful end device across all computing nodes imposes a performance bottleneck. Meanwhile, coupled heterogeneity in data distributions and resource capabilities across tiers disrupts hierarchical knowledge transfer, leading to biased updates and degraded performance. Furthermore, the mobility and fluctuating connectivity of computing nodes in EECC environments introduce complexities in dynamic node migration, further compromising the robustness of the training process.}
\textcolor{black}{To address multiple challenges within a unified framework}, we propose End-Edge-Cloud Federated Learning with Self-Rectified Knowledge Agglomeration (FedEEC), which is a novel EECC-empowered FL framework that allows the trained models from end, edge, to cloud to grow larger in size and stronger in generalization ability. FedEEC introduces two key innovations: (1) Bridge Sample Based Online Distillation Protocol (BSBODP), which enables knowledge transfer between neighboring nodes through generated bridge samples, and (2) Self-Knowledge Rectification (SKR), which refines the transferred knowledge to prevent suboptimal cloud model optimization. The proposed framework effectively handles both cross-tier resource heterogeneity and effective knowledge transfer between neighboring nodes, \textcolor{black}{while satisfying the migration-resilient requirements of EECC.} Extensive experiments on three datasets demonstrate that FedEEC achieves significantly higher cloud model accuracy compared with state-of-the-art methods.
\end{abstract}
}
\maketitle

\IEEEdisplaynontitleabstractindextext

%

\IEEEpeerreviewmaketitle
\vspace{20pt}
\IEEEraisesectionheading{\section{Introduction}}
\IEEEPARstart{W}{ith} the growing boom of Internet of Things (IoT), 6G, and Artificial Intelligence (AI), the proliferation of intelligent devices is accelerating at an exponential rate. According to the predictions from Statista, the number of IoT connections worldwide will reach 32.1 billion by 2030 \cite{iot-predict}. 
This surge is accompanied by an unprecedented volume of data generated at the network end, offering immense potential for training advanced AI models and enhancing user experiences across a wide range of applications \cite{baccour2022pervasive}. However, concerns over data privacy and ownership make it impractical to conduct centralized end-side data processing on the cloud, resulting in isolated data islands \cite{yang2019federated} that impede the development of comprehensive AI models and large-scale data-driven insights. To address this challenge, Federated Learning (FL) has emerged as a promising solution within the realm of Privacy Computing (PC), allowing multiple devices to jointly train machine learning models without sharing on-device data with the cloud \cite{nguyen2021federated}. In conventional FL \cite{mcmahan2017communication,li2020federated,karimireddy2020scaffold}, the server and end devices (clients) collaborate through iterative rounds of model parameters aggregation and distribution, effectively bridging isolated data islands while maintaining privacy guarantees.

\textcolor{black}{Alongside the evolution of FL,} the computing paradigm is shifting from centralized cloud-based processing to End-Edge-Cloud Collaboration (EECC) \cite{duan2022distributed}, where computational tasks are distributed across cloud servers, edge servers, and end devices.
This paradigm offers reduced network latency, lower bandwidth consumption, and enhanced system reliability \cite{wang2024end}. FL can benefit from the EECC paradigm to fully exploit pervasive devices and rapidly access large amounts of end-side data \cite{bao2022federated}. 
However, prevailing FL methods \cite{mcmahan2017communication,li2020federated,karimireddy2020scaffold} adopt a simple client-server architecture with two tiers, where the server is either on the cloud or the edge. \textcolor{black}{Such methods lack the architectural flexibility to fully exploit the potential of EECC environments, which require multi-tier collaboration across end devices, edge nodes, and cloud servers.}

To overcome this limitation, Hierarchical Federated Learning (HFL) \cite{liu2020client} is proposed to extend conventional FL by enabling multi-tier model aggregation across different tiers of computing nodes, unlocking the potential for efficient collaboration in an EECC system. 
\textcolor{black}{Despite this promising potential, fully realizing the benefits of HFL in EECC environments is far from straightforward. The heterogeneous and dynamic nature of an End-Edge-Cloud Network (EEC-NET) introduces several interrelated challenges that hinder collaborative training across tiers.}

\textcolor{black}{The first challenge arises from computing power heterogeneity across end, edge, and cloud nodes.} 
Specifically, cloud servers are equipped with powerful computing capabilities and storage resources, which outperform edge servers, and end devices possess relatively constrained resources \cite{tak2020federated}. Current HFL methods \cite{liu2020client,wang2023accelerating,yang2023hierarchical,liu2022hierarchical} require imposing the same model structure on all computing nodes for model aggregation, inevitably limiting the scale of trained models by the least powerful end devices due to the bottleneck effect and causing resource under-utilization over edge and cloud computing nodes. 
\textcolor{black}{The second challenge arises from the necessity for effective model update transfer from lower- to upper-tier nodes.} In the end-edge-cloud hierarchy, the model updates learned by lower-tier devices must be effectively transferred upwards to ensure the performance of the cloud model. \textcolor{black}{Due to the variations in data distributions and resource constraints across tiers, the model updates learned at lower tiers might be biased or incomplete, potentially introducing misleading patterns that propagate upward through the network hierarchy, ultimately resulting in suboptimal optimization of the cloud model.} The third challenge arises from the introduction of the end-edge-cloud hierarchy, where computing nodes may need to switch between different upper-tier servers (called dynamic node migration) due to factors such as mobility or unstable connections \cite{wang2019dynamic,jsac-dynamic}. Some HFL methods \cite{deng2023hierarchical,wang2023accelerating}, with their static coordination algorithms, are unable to handle such dynamic node associations, \textcolor{black}{resulting in reduced robustness in real-world deployments.}

To effectively address the aforementioned challenges posed by the heterogeneous and dynamic nature of EECC, several key issues should be considered to guide the design of HFL. First, model sizes must be adapted to match the computational capabilities of nodes at different tiers, allowing resource-rich edge and cloud nodes to host larger models without being constrained by the capabilities of end devices. Second, \textcolor{black}{the quality of exchanged information} between neighboring nodes must be optimized, ensuring that every \textcolor{black}{model update} transferred from lower-tier nodes contributes to the enhancement of models at higher tiers rather than degrades their performance. \textcolor{black}{Finally, it is crucial to enable seamless interaction between nodes with arbitrary model structures, ensuring compatibility with the dynamic migration of computing nodes commonly observed in EECC environments.}


To this end, we propose a novel HFL framework suitable for EECC paradigm, named End-Edge-Cloud Federated Learning with Self-Rectified Knowledge Agglomeration (FedEEC). FedEEC is designed to enable recursive collaboration across computing nodes from end devices to edge and cloud servers, with model size and generalization ability increasing progressively from lower-tier to higher-tier nodes. Specifically, FedEEC introduces two core innovations: Bridge Sample-Based Online Distillation Protocol (BSBODP) and Self-Knowledge Rectification (SKR). BSBODP tackles the challenge of heterogeneous model collaboration by facilitating model-agnostic \cite{makd} knowledge transfer between neighboring nodes in an EEC-NET with a tree-topology. This is achieved through a pre-trained lightweight decoder that generates synthetic samples (referred to as bridge samples), which are used to extract logits for knowledge distillation across neighboring nodes. 
SKR mitigates the propagation of misleading knowledge by refining the transferred predictions through a series of corrective steps. It employs knowledge queues to cache reliable historical predictions, applies maximum likelihood estimation to adjust class probabilities based on this cached knowledge, and uses relative entropy minimization to preserve the inter-class relationships of the corrected probabilities. We also demonstrate that FedEEC supports arbitrary non-cloud nodes switching between upper-tier servers, providing flexibility guarantees for knowledge agglomeration in multi-tier networks.

In summary, the main contributions of this paper are as follows:
\begin{itemize}
    \item 
    We propose a novel HFL framework suitable for EECC paradigm, named End-Edge-Cloud Federated Learning with Self-Rectified Knowledge Agglomeration (FedEEC), which recursively organizes computing nodes via a customized interaction protocol and enables models trained on the end, edge, and cloud nodes to grow larger in size and stronger in generalization ability.
    \item 
    \textcolor{black}{We introduce Bridge Sample Based Online Distillation Protocol (BSBODP) and Self-Knowledge Rectification (SKR) to facilitate model-agnostic collaborative training among neighboring nodes, and to refine the transferred class probabilities to mitigate the propagation of misleading predictions, respectively. Both mechanisms are theoretically grounded to support the dynamic migration of computing nodes in EECC environments.}
    \item 
    We validate the effectiveness of FedEEC in an end-edge-cloud architecture over SVHN, CIFAR-10, and CINIC-10 datasets. Empirical results demonstrate that FedECC achieves better cloud model accuracy in all considered settings compared with state-of-the-art methods.
\end{itemize}

\section{Preliminary}
\subsection{End-Edge-Cloud Collaboration}
  \label{eecc}
    End-Edge-Cloud Collaboration (EECC) is an emerging computing paradigm that coordinates multi-tier heterogeneous computing nodes including central cloud servers, numerous edge servers, and massively distributed end devices, aiming to work on large-scale computing tasks collaboratively. Typically, end devices are located at the periphery of the network, where they generate data and perform data-intensive AI applications.
    Edge servers bridge to connect end devices and cloud servers, distributed at intermediate locations along the network. Cloud servers, situated at the network's core, are equipped with sufficient computational and storage resources, facilitating the coordination of underlying end devices and edge servers.

\begin{figure}[t]
	\centering
	\includegraphics[width=0.5\textwidth]{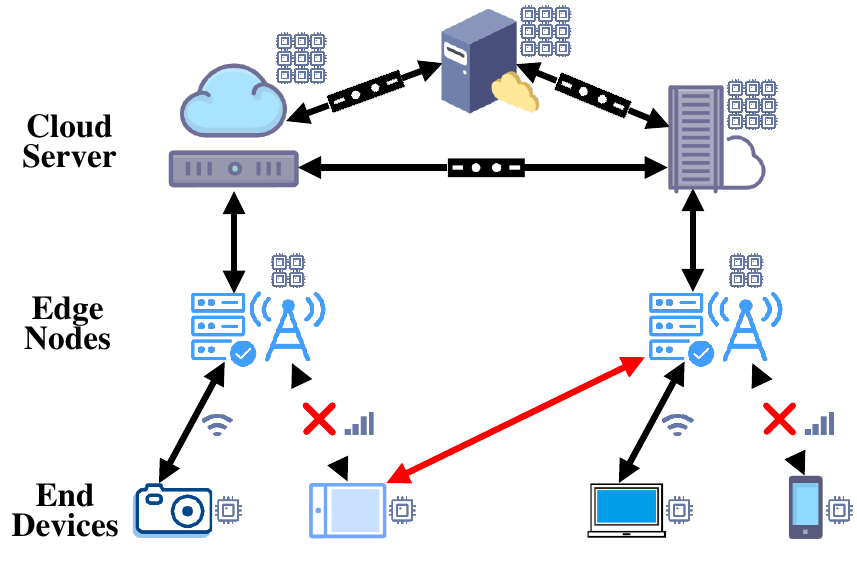}
	\caption{\textcolor{black}{Hierarchical topology of end-edge-cloud Collaboration with dynamic node migration.}}
	\label{framework}
\end{figure}

    In this paper, we consider a typical End-Edge-Cloud Network (EEC-NET) with a tree topology (shown in Fig. \ref{framework}) denoted as $G=(\mathcal{V},\mathcal{E})$, where $\mathcal{V}$ is the set of computing nodes and $\mathcal{E}$ is the set of communication links.
    The EEC-NET $G$ is structured with a single root node $r\in \mathcal{V}$ and one or more leaf nodes, which form the set $\mathcal{L} \subset \mathcal{V}$. 
    Each computing node $v\in \mathcal{V}$ except leaf nodes has one or more children, collectively forming a set $Child(v)$. 
    Any computing node $v\in \mathcal{V}$ except the root node is associated with a single parent node, denoted as $Parent(v)$. 
    Additionally, we define $Leaf(v)$ as the set of all leaf nodes within the sub-tree that seeks the computing node $v$ as its root, and hence we have $Leaf(r)=\mathcal{L}$.   
    For organization, the computing nodes in the EEC-NET are arranged into hierarchical tiers. Assuming the EEC-NET $G$ have a total of $T$ tiers, and the computing nodes in tier $t\in\{1,2,...,T\}$ form a set $\mathcal{V}_t$, so that we have $\mathcal{V}_T=\mathcal{L}$ and $\mathcal{V}_1=\{r\}$.
    For reasons of flexibility, EECC should take into account the dynamic migration of computing nodes, as illustrated in Fig. \ref{framework}. Each non-root node in the EEC-NET should dynamically reestablish its parent node to efficiently redistribute workloads and recover from disruptions, ensuring uninterrupted operations and improving overall system functionality.

    \subsection{Hierarchical Federated Learning Empowered by End-Edge-Cloud Collaboration}
    We consider a scenario where $K=|\mathcal{L}|$ end devices (called clients) participate in Hierarchical Federated Learning (HFL) over an EEC-NET. Each client $k\in\{1,2,...,K\}$ owns a local dataset ${\mathcal{D}^k} = \bigcup\limits_{i = 1}^{{N^k}} {\{ (X_i^k,y_i^k)\} }$, in which $N^k$ is the number of samples in $\mathcal{D}^k$, and $X_i^k$, $y_i^k$ are the data and label of the $i$-th sample in $\mathcal{D}^k$, respectively.
    Besides, each node $v\in \mathcal{V}$ holds a model $Model(v)$ with parameters $W^v$, and can exchange information with its parent and child nodes, i.e., $\{Parent(v)\} \cup Child(v)$.
    Our goal is to obtain the model in the root node $Model(r)$ that minimizes its training loss $L_{train}(\cdot)$ over all private data on devices, formulated as:
	\begin{equation}
		\mathop {{\rm{argmin}}}\limits_{{W^r}} {L_{train}}(\bigcup\limits_{k = 1}^K {{\mathcal{D}^k}} ;{W^r}).
	\end{equation}

\begin{figure}[t]
	\centering
    \includegraphics[width=0.35\textwidth]{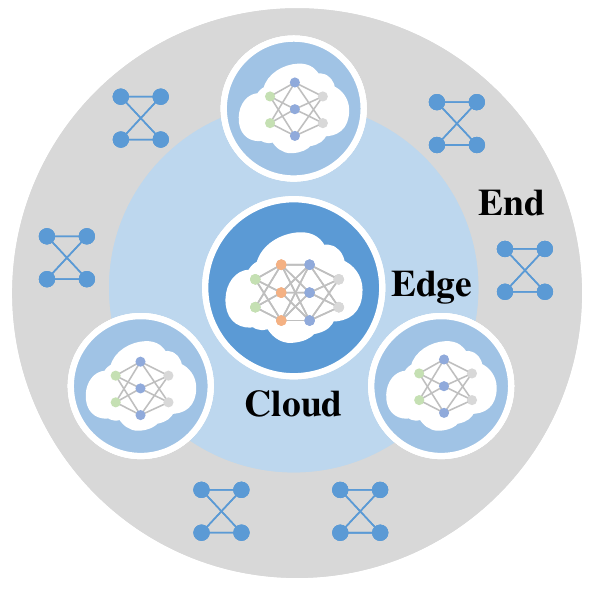}
	\caption{\textcolor{black}{Expected model scales distribution across tiers in an end-edge-cloud network.}}
	\label{model}
\end{figure}

    To fully exploit heterogeneous resources and accommodate the differentiated capabilities of end, edge, and cloud computing nodes, it is crucial to guarantee edge and cloud servers, which possess more powerful computational resources, deploy larger models than resource-constrained end devices, as shown in Fig. \ref{model}. These larger models are more capable of extracting complex and generalized patterns from private data, thereby fully leveraging upper-tier resources to deliver substantial improvements in system performance.

\subsection{Interaction Protocols in Federated Learning}
Interaction protocols in FL define the rules for information exchange between computing nodes.
Prevailing FL interaction protocols are based on parameters interaction like FedAvg \cite{mcmahan2017communication}, where upper-tier nodes aggregate and distribute model parameters between themselves and their subordinate nodes.
Without considering child selection, this aggregation protocol typically keeps to the following steps to update the model on the non-root computing node $v^{*}$ and its parent node $Parent(v^*)$:
	\begin{itemize}
		\item $v^{*}$ uploads its model parameters $W^{v^{*}}$ to $Parent(v^*)$.
		\item $Parent(v^*)$ waits for all its child nodes $Child(Parent(v^*))$ to finish uploading their model parameters.
		\item $Parent(v^*)$ aggregates the uploaded model parameters according to the following assignment to obtain updated model parameters:
		\begin{equation}
			{W^{Parent({v^*})}} \leftarrow \frac{{\sum\limits_{v \in Child(Parent({v^*}))} {{W^v} \cdot \sum\limits_{u \in Leaf(v)} {|{\mathcal{D}^u}|} } }}{{\sum\limits_{v \in Child(Parent({v^*}))} {\sum\limits_{u \in Leaf(v)} {|{\mathcal{D}^u}|} } }}
		\end{equation}
		where ${|{\mathcal{D}^u}|}$ denotes the number of local data samples from the device corresponding to the leaf node $u$.
		\item $Parent(v^*)$ distributes the updated model parameters ${W^{Parent({v^*})}}$ to all its child nodes.
		\item $v^{*}$ updates its model parameters with those downloaded from its parent node $Parent(v^*)$.
	\end{itemize}




\begin{table*}[]
\centering
\caption{Comparison of FedEEC with existing FL methods.}
\setlength{\tabcolsep}{10pt}
\label{compareison-fedeec}
\renewcommand{\arraystretch}{0.85}
\begin{tabular}{l|c|c|c|c}
\hline
\multicolumn{1}{c|}{\textbf{Method}} & \textbf{\begin{tabular}[c]{@{}c@{}}End-Edge-Cloud\\ Collaboration\end{tabular}} & \textbf{\begin{tabular}[c]{@{}c@{}}Larger Model\\ on the Cloud\end{tabular}} & \textbf{\begin{tabular}[c]{@{}c@{}}Dynamic\\ Migration\end{tabular}} & \textbf{\begin{tabular}[c]{@{}c@{}}Training without\\ Public Data\end{tabular}} \\ 

\hline

FedAvg \cite{mcmahan2017communication}                               & \ding{55}                                                                              & \ding{55}                                                                           & \ding{55}                                                                   & \ding{51}                                                                             \\
FedProx \cite{li2020federated}                             & \ding{55}                                                                              & \ding{55}                                                                           & \ding{55}                                                                   & \ding{51}                                                                             \\
FedGems \cite{cheng2021fedgems}                              & \ding{55}                                                                              & \ding{51}                                                                          & \ding{55}                                                                   & \ding{55}                                                                              \\
FedET \cite{cho2022heterogeneous}                                & \ding{55}                                                                              & \ding{51}                                                                          & \ding{55}                                                                   & \ding{55}                                                                              \\
FedRolex \cite{alam2022fedrolex}                            & \ding{55}                                                                              & \ding{51}                                                                          & \ding{55}                                                                   & \ding{51}                                                                             \\
FedGKT \cite{he2020group}                              & \ding{55}                                                                              & \ding{51}                                                                          & \ding{55} & \ding{51} 
\\
 
FedDKC \cite{wu2022exploring}                              & \ding{55}                                                                              & \ding{51}                                                                          & \ding{55}                                                                   &
\ding{51}                                                                             \\
DemLearn \cite{nguyen2023self}                            & \ding{51}                                                                             & \ding{55}                                                                           & \ding{55}                                                                   & \ding{51}                                                                             \\
FedHKT \cite{deng2023hierarchical}                              & \ding{51}                                                                             & \ding{51}                                                                          & \ding{55}                                                                   & \ding{55}                                                                              \\
HierFAVG \cite{liu2020client}                            & \ding{51}                                                                             & \ding{55}                                                                           & \ding{51}                                                                  & \ding{51}                                                                             \\
FedCH \cite{wang2023accelerating}                             & \ding{51}                                                                             & \ding{55}                                                                           & \ding{55}                                                                  & \ding{51}                                                                             \\
HierMo \cite{yang2023hierarchical}                              & \ding{51}                                                                             & \ding{55}                                                                           & \ding{51}                                                                  & \ding{51}                                                                             \\ \hline
\textbf{FedAgg \cite{wu2024agglomerative}}               & \ding{51}                                                                             & \ding{51}                                                                          & \ding{51}                                                                  & \ding{51}                                                                             \\
\textbf{FedEEC}              & \ding{51}                                                                             & \ding{51}                                                                          & \ding{51}                                                                  & \ding{51}                                                                             \\ 
\hline

\end{tabular}
\end{table*}

\section{Motivation}
\subsection{Knowledge Agglomeration}
To exploit the computational power of edge and cloud nodes within the HFL process, we prompt the intuitive motivation of training larger models on these computing nodes than end devices for achieving higher accuracy with stronger generalization abilities. However, a key challenge arises: how can we effectively facilitate collaboration between models with different structures across the hierarchical tiers of an EEC-NET? Traditional parameters aggregation, as used in prevailing EECC-empowered FL methods \cite{liu2020client,wang2022accelerating,yang2023hierarchical}, is inadequate because it requires uniform model structures to be adopted across all computing nodes, limiting the scale of trained models by the least powerful end devices and thus under-utilizing the computational power of edge and cloud servers.

To overcome this challenge, we propose utilizing online distillation \cite{anil2018large,wang2021knowledge} as an interaction protocol among computing nodes at different tiers, which enables different nodes to train models with heterogeneous structures via iteratively exchanging the logits (also called knowledge) between two neighboring nodes \textcolor{black}{(also called parent-child node pairs)}, guiding reciprocal model training while supporting potential knowledge agglomeration propagating up the tree hierarchy toward the cloud computing node. 
However, directly integrating online distillation into EECC-powered FL may raise privacy concerns about sharing data on devices, as the same sample is required to extract logits across different computing nodes. Therefore, we propose to employ synthetic data as a bridge to transfer knowledge across multi-tier nodes with the assistance of a lightweight pre-trained autoencoder. This synthetic data is generated by the decoder using either encoded or received embeddings, thereby enabling the computation of logits without revealing raw data on devices. Empowered by this interaction protocol, our hierarchical knowledge transfer framework allows the FL system to take full advantage of the heterogeneous computing capabilities across different tiers in EEC-NET, from resource-constrained end devices to more powerful edge and cloud nodes.

\subsection{Knowledge Rectification}
The bottom-up knowledge agglomeration process in EECC-empowered FL faces challenges in ensuring the reliability of transferred knowledge between neighboring nodes. Due to the resource constraints of end devices and the inherently scarce, non-IID nature of their data, models trained at these lower tiers may generate biased or misleading predictions. \textcolor{black}{When erroneous knowledge propagates upwards to edge or cloud nodes, it amplifies errors and ultimately degrades the optimization of the cloud model.}


To tackle this issue, we adaptively rectify knowledge within every computing node to mitigate the propagation of misleading patterns during interactions. Specifically, we leverage historical knowledge queues and statistical probabilistic adjustments to correct potentially misleading predictions, guaranteeing precise adjustment of softmax probabilities for label-relevant classes, while maintaining the similarity integrity for the remaining classes with theoretical support. This adaptive rectification mechanism not only improves the quality of transferred knowledge but also enhances the robustness of the entire EECC-empowered FL system against potential biases and errors introduced by individual nodes.

\begin{figure*}[t]
	\centering
	\includegraphics[width=1.0\textwidth]{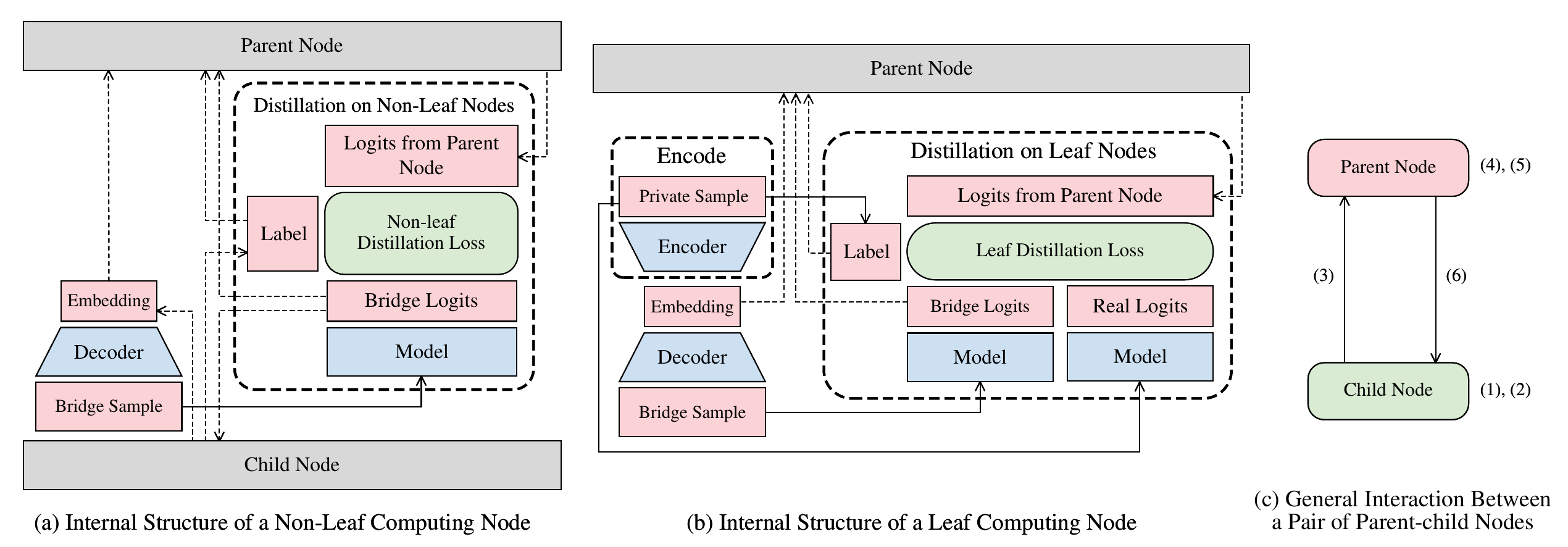}
	\caption{Overview of BSBODP. 
		(1) Child node distillation on bridge samples.
		(2) Logits extraction on child node.
		(3) Upload logits to parent node.
		(4) Parent node distillation on bridge samples. 
		(5) Logits extraction on parent node.
		(6) Distribute logits to child node.}
	\label{bsodp}
\end{figure*}

\begin{figure}[t]
	\centering
	\includegraphics[width=0.45\textwidth]{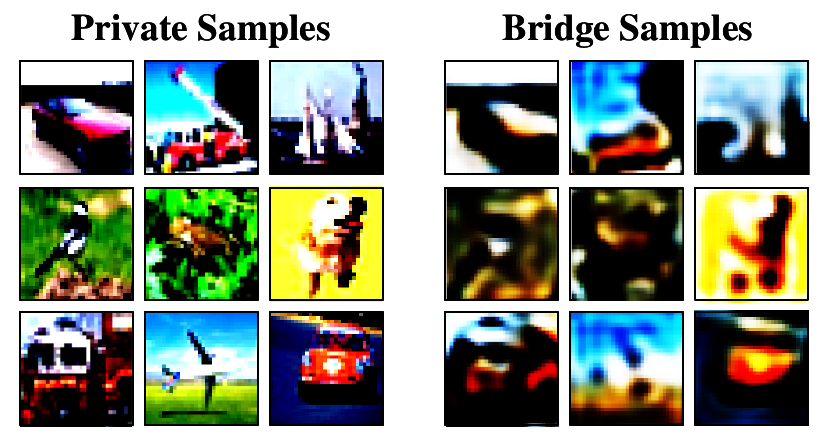}
	\caption{Comparison of private samples and bridge samples.}
	\label{bridge}
\end{figure}

\section{End-Edge-Cloud Federated Learning with Self-Rectified Knowledge Agglomeration}
\subsection{Framework Overview}
We introduce End-Edge-Cloud Federated Learning with Self-Rectified Knowledge Agglomeration (FedEEC), a novel EECC-empowered FL framework that supports tier-by-tier model scaling across end, edge, and cloud computing nodes within an EEC-NET. A detailed comparison of FedEEC with existing FL methods is provided in Table \ref{compareison-fedeec}. In FedEEC, each node can adopt a model structure that matches its own computing power, and the lower-tier nodes can iteratively transfer their learned knowledge to the upper-tier nodes via a customized interaction protocol. \textcolor{black}{Therefore, the cloud model can ultimately integrate the knowledge from all nodes and capitalize on its inherently larger capacity to capture complex patterns, thereby delivering superior performance.} To facilitate general interaction between computing nodes in FedEEC, we propose the Bridged Sample-Based Online Distillation Protocol (BSBODP), which ensures privacy and flexibility during multi-tier collaborative training by performing logit-based mutual knowledge distillation on generated synthetic samples. Furthermore, to enhance the quality of knowledge transfer and improve the robustness of the overall FL system, we design a Self-Knowledge Rectification (SKR) mechanism. Leveraging principles of maximum likelihood estimation and relative entropy minimization, SKR dynamically rectifies the transferred softmax probabilities based on reliable historical classification results, enhancing resistance to the propagation of misleading knowledge during interactions. By combining the advantages of both BSBODP and SKR, FedEEC recursively organizes computing nodes of each tier in EEC-NET, while complying with continuous growth in model scale and generalization ability from the bottom to the top. Additionally, FedEEC can also handle scenarios where any non-root computing node dynamically changes its
parents, which ensures deployment flexibility in a realistic EEC-NET. In the following subsections, we will provide detailed elaborations on the technical details of FedEEC and demonstrate its deployment flexibility within EEC-NET with theoretical support.


\subsection{Bridge Sample Based Online Distillation Protocol}
Fig. \ref{bsodp} provides the overview of our designed BSBODP, where each computing node is equipped with a pre-trained decoder and a conventional complete model (the model to be trained on the computing node). In particular, each leaf computing node additionally preserves a pre-trained encoder (forms a pre-trained autoencoder with the aforementioned decoder) for encoding local data into embeddings, which will be transmitted to its parent node until reaching the root node.
During execution, each computing node utilizes the decoder to generate bridge samples that match the data distribution of its corresponding leaf nodes based on the embeddings encoded from private data or uploaded by sub-nodes.
These bridge samples are used as intermediaries, facilitating online distillation between every pair of parent-child computing nodes in the tree hierarchy, with alignment ensured by the same embedding. As illustrated in Fig. \ref{bridge}, it is challenging to recover raw information from the embeddings of private samples since the embeddings are generated based on an extremely lightweight autoencoder ($<$50K model parameters) that are pre-trained on a super large open dataset (such as ImageNet \cite{deng2009imagenet}).
The autoencoder used to generate bridge samples is intentionally designed with limited capacity, making it hard to reconstruct fine-grained features of the original data. This makes online distillation a privacy-acceptable interaction protocol between neighboring computing nodes.

\begin{algorithm}[t]
	\caption{Bridge Sample Based Online Distillation Protocol (BSBODP)}
	\SetAlgoNoLine

	\label{alg1}
	\begin{spacing}{0.65}
        
		\KwIn{$v^1$,$v^2$}
		\KwOut{Trained $Model(v^1)$, $Model(v^2)$}
        \begin{algorithmic}[0]
            
            \Procedure{BSBODP}{$v^1$, $v^2$}
           \State \Call{BSBODP-Directional}{$v^1$, $v^2$}
          \State \Call{BSBODP-Directional}{$v^2$, $v^1$}
          \State \Return{$Model(v^1)$, $Model(v^2)$}
		\EndProcedure

           \Procedure{BSBODP-Directional}{$v^{S}$, $v^{T}$}
            \State 
                $v^{T}$ generates bridge samples $dec(\varepsilon)$ from all stored embeddings $\varepsilon$
            \State $v^{T}$ extracts logits ${z^\varepsilon} = f(dec(\varepsilon );{W^{{v^T}}})$ on bridge samples, and transmits the results to $v^{S}$
            \State \If{$v^{S} \notin \mathcal{L}$}
            {
            \State Optimize $W^{v^{S}}$ according to Eq. (\ref{optim-non-leaf})
            }
            \State \Else
            {
            \State Optimize $W^{v^{S}}$ according to Eq. (\ref{optim-leaf})
            }
            \EndProcedure 
           \end{algorithmic}
	\end{spacing}
\end{algorithm}

We formulate the process of BSBODP as follows: the decoder on each computing node is defined as $dec(\cdot)$, and the encoder on each leaf computing node is defined as $enc(\cdot)$. 
In addition, the model to be trained on any computing node $Model(v^*),v^* \in \mathcal{V}$ can give an inference on the input data via $f(\cdot;W^{v^*})$.
When performing upward knowledge distillation, the child node in any pair of parent-child computing nodes acts as the teacher $v^T$, and the parent node acts as the student $v^S$. When performing downward knowledge distillation, the roles are reversed.
Specifically, each model on non-leaf student computing node $Model(v^{S}),\forall v^{S}\notin \mathcal{L}$ distills the knowledge from the model on its teacher $Model(v^{T})$ over the bridge samples that are generated from the embeddings $\varepsilon$ corresponding to $
\bigcup\limits_{u \in Leaf({v^{S}})} {{{\cal D}^u}}  \cap \bigcup\limits_{u \in Leaf({v^{T}})} {{{\cal D}^u}}$.
With generated bridge samples as intermediates, $Model(v^{S})$ is optimized according to the non-leaf distillation loss as follows:
\begin{equation}
	\begin{array}{l}
	\; \; \; \; \mathop {\min }\limits_{{W^{{v^{S}}}}} {L_{non - leaf}}\\ 
	= \mathop {\min }\limits_{{W^{{v^{S}}}}} [{L_{CE}}(\tau (f(dec(\varepsilon );{W^{{v^{S}}}})),{y^\varepsilon }) + \\
	\; \; \; \; \beta  \cdot KL(\tau (f(dec(\varepsilon );{W^{{v^{S}}}}))||\tau (\frac{{{z^\varepsilon }}}{T}))]\\
	 = \mathop {\min }\limits_{{W^{{v^{S}}}}} [{L_{CE}}(\tau (f(dec(\varepsilon );{W^{{v^{S}}}})),{y^\varepsilon }) +
	 \\ \; \; \; \; \beta  \cdot KL(\tau (f(dec(\varepsilon );{W^{{v^{S}}}}))||\tau (\underbrace {\frac{{f(dec(\varepsilon );{W^{{v^{T}}}})}}{T}}_{\frac{{{z^\varepsilon }}}{T}}))],
	\end{array}
	\label{optim-non-leaf}
\end{equation}
where $L_{CE}(\cdot)$ is the cross-entropy loss function, $KL(\cdot)$ is the Kullback-Leibler divergence loss function, and $\beta$ is the distillation weight.
Moreover, $y^\varepsilon$ and $z^\varepsilon$ are the labels and extracted logits of bridge samples corresponding to embeddings $\varepsilon$ uploaded from child nodes, and $\varepsilon$ is ultimately generated at the leaf nodes according to the following equation:
\begin{equation}
    \varepsilon  = enc({X^*}), \forall ({X^*},{y^*}) \in \bigcup\limits_{u \in Leaf({v^S})} {{\mathcal{D}^u}}  \cap \bigcup\limits_{u \in Leaf({v^T})} {{\mathcal{D}^u}}.
    \label{logits-non-leaf}
\end{equation}
Moreover, each model on leaf student computing node $Model(v^{S}),\forall v^{S}\in \mathcal{L}$ is optimized subject to a linear combination of the non-leaf distillation loss over generated bridge samples and the local training losses (controlled by a weight $\gamma$) over private samples $(X^*,y^*)\in \mathcal{D}^{v^{S}}$, that is:
\begin{equation}
		\begin{array}{l}
			 \; \; \; \; \mathop {\min }\limits_{{W^{{v^{S}}}}} {L_{leaf}}\\
			 = \mathop {\min }\limits_{{W^{{v^{S}}}}} {L_{CE}}(f({X^*};{W^{{v^{S}}}}),{y^*}) + \gamma  \cdot {L_{non - leaf}},
		\end{array}
	\label{optim-leaf}
\end{equation}
where
\begin{equation}
	{\quad \varepsilon  = enc({X^*})}, \forall (X^*,y^*)\in \mathcal{D}^{v^{S}}.
    \label{logits-leaf}
\end{equation}
Once the above steps are completed, $v^{S}$ and $v^{T}$ need to swap their roles and optimize in the opposite direction following the above constraints, as shown in Algorithm \ref{alg1}.
Thereout, the models to be trained on every pair of parent-child computing nodes can learn from each other, and the representation learned through knowledge distillation can be propagated to the model on the cloud tier by tier, starting from the leaf nodes.

\color{black}
\subsection{Self-Knowledge Rectification}
To ensure reliable knowledge transfer in BSBODP, we propose Self-Knowledge Rectification (SKR), a mechanism designed to adaptively rectify misleading attributions in bridge sample knowledge caused by resource limitations and non-IID data at lower-tier nodes, preventing error propagation and enhancing system performance. SKR introduces a rectification process at each computing node, evaluating the predicted probabilities to identify potential inaccuracies and correcting them before transferring them to neighboring nodes. This process ensures that the transferred knowledge is more reliable and aligned with well-attributed historical predictions. In the following subsections, we detail the procedure of SKR and demonstrate how it systematically enhances hierarchical knowledge transfer within the EEC-NET.

\subsubsection{Misleading Class Probabilities Identification}
\textcolor{black}{Consider an arbitrary computing node $v^* \in \mathcal{V}$ in the EEC-NET, we consider the cases where an arbitrary bridge sample $dec(\varepsilon)$ belonging to class $c=y^*$ is used to extract logits $z^\varepsilon$. For a specific $dec(\varepsilon)$, the original probability estimation for class $c$ is derived from the softmax-processed logits, that is:}
\begin{equation}
    p_c=\tau (\frac{z^\varepsilon}{T})_c=\tau (\frac{{f(dec(\varepsilon );{W^{{v^*}}})}}{T})_c,
\end{equation}
where $p_c$ is the predicted probability of classifying $dec(\varepsilon)$ as class $c$. 
However, the model’s prediction confidence may be misplaced, resulting in the estimated $c$-class probability $p_c$ being incorrectly attributed. Specifically, for a bridge sample of class $c$, \textcolor{black}{the model might assign higher probability estimates to one or more non-c classes than to class $c$}, that is:
\begin{equation}
    {p_c} < \tau (\frac{z^\varepsilon}{T})_i,\exists i \in \{ 1,2,...,C\} \wedge i \neq c .
\end{equation}
\color{black}{This condition highlights scenarios where the class predictions are incorrectly attributed, which can introduce biased updates that propagate through the EEC-NET hierarchy, ultimately degrading the performance of the cloud-based model.}

\subsubsection{Knowledge Queues}
To assist the knowledge rectification process, we introduce a knowledge queue to store historical class probabilities from instances where the model on $v^*$ correctly estimates $c$-class bridge samples. The knowledge queue for class $c$ can hold up to $B$ elements and is denoted as $\mathcal{B}_c =[{q_c^1, q_c^2, ..., q_c^B}]$, where each element in the queue $q_c^j \in \mathcal{B}_c, \forall j \in \{1, 2, ..., B\}$ records the value of $p_c$ from instances where the model correctly identified any $c$-class bridge sample $dec(\varepsilon)$ as belonging to class $c$ in previous inference processes. When the queue reaches its capacity, the addition of new elements results in the removal of the oldest elements, ensuring that $\mathcal{B}_c$ always contains the most recent correctly attributed knowledge.

\subsubsection{Knowledge Rectification for Label-relevant Classes}
To rectify knowledge of label-relevant classes, we utilize historical knowledge stores in knowledge queues and apply maximum likelihood estimation to compute rectified probability estimates $p_c'$ across bridge samples $dec(\varepsilon)$ belonging to class $c$. Assuming the historical well-attributed probabilities $\{q_c^1, q_c^2, ..., q_c^B\}$ are independently and identically distributed, following a certain probability distribution $P(\mathcal{B}_c|p_c')$. We define the likelihood function $L_{p_c'}$ as the joint probability of $\{q_c^1, q_c^2, ..., q_c^B\}$ conditioned on the parameter $p_c'$, that is:
\begin{equation}
    L_{p_c'}=P(q_c^1,q_c^2,...,q_c^B|p_c').
\end{equation}
Given that $\{q_c^1, q_c^2, ..., q_c^B\}$ are mutually independent, we can derive:
\begin{equation}
    L_{p_c'} =P(q_c^1,q_c^2,...,q_c^B|p_c')= \prod_{i=1}^B P(q_c^i | p_c').
\end{equation}
Therefore, the log-likelihood function $\ell(p_c')$ can be formulated as:
\begin{equation}
    \ell_{p_c'} = \log L_{p_c'} = \log \left( \prod_{i=1}^B P(q_c^i | p_c') \right) = \sum_{i=1}^B \log P(q_c^i | p_c').
\end{equation}
Assuming $q_c^i \in \mathcal{B}_c, \forall i \in \{1, 2, ..., B\}$ follows a Gaussian distribution with mean $p_c'$ and variance $s^2$, i.e., $q_c^i \sim \mathcal{N}(p_c', s^2), \forall i \in \{1, 2, ..., B\}$. Then, the probability distribution $P(q_c^i | p_c')$ can be formulated as:
\begin{equation}
    P(q_c^i | p_c') = \frac{1}{\sqrt{2 \pi }s} \exp \left( -\frac{(q_c^i - p_c')^2}{2 s^2} \right).
    \label{distribution}
\end{equation}
Substituting Eq. (\ref{distribution}) into Eq. (\ref{log-likely}), we have:
\begin{equation}
\begin{array}{l}
\ell_{p_c'}
 = \sum\limits_{i = 1}^B {\log } \left( {\frac{1}{{\sqrt {2\pi } s}}\exp \left( { - \frac{{{{(q_c^i - {p_c'})}^2}}}{{2{s^2}}}} \right)} \right)\\
\;\;\;\; \; \; = \sum\limits_{i = 1}^B {\left( {\log \frac{1}{{\sqrt {2\pi } s}} - \frac{{{{(q_c^i - {p_c'})}^2}}}{{2{s^2}}}} \right)} \\
 \;\;\;\; \; \;  = B\log \frac{1}{{\sqrt {2\pi } s}} - \frac{1}{{2{s^2}}}\sum\limits_{i = 1}^B {{{(q_c^i - {p_c'})}^2}}.
\end{array}
 \label{log-likely}
\end{equation}
To maximize the log-likelihood function $\ell(p_c')$, we compute the derivative with respect to $p_c'$ in Eq. (\ref{log-likely}) and set it to zero, which means:
\begin{equation}
    \frac{d\ell_{p_c'}}{dp_c'} = - \frac{1}{2 s^2} \sum_{i=1}^B 2 (q_c^i - p_c') (-1) = \frac{1}{s^2} \sum_{i=1}^B (q_c^i - p_c') = 0.
\end{equation}
Solving the above equation, we can derive the result of $p_c'$, that is:
\begin{equation}
    p_c' = \frac{1}{B} \sum_{i=1}^B q_c^i.
    \label{pc'}
\end{equation}

\subsubsection{Knowledge Rectification for Non-label Classes}
To rectify knowledge of non-label classes, we aim to preserve their relative relationships while accommodating the rectified probability of label-relevant classes. Specifically, we reconstruct the transferred estimated probabilities to $\mathcal{Q}={[p_1',p_2',…,p_C']}$, ensuring that the original predicted probabilities $\mathcal{P}={[p_1,p_2,…,p_C]}$ have the smallest relative entropy (Kullback-Leibler divergence) with $\mathcal{Q}$, so as to preserve the original class relevance information as much as possible. The expanded form of relative entropy between $\mathcal{P}$ and $\mathcal{Q}$ is as follows:
\begin{equation}
    {KL}(\mathcal{P} \parallel \mathcal{Q}) = \sum_{i=1}^C p_i \log \frac{p_i}{p_i'}.
\end{equation}
We aim to minimize the relative entropy between $\mathcal{P}$ and $\mathcal{Q}$ under the premise of Eq. (\ref{pc'}) and the basic properties of probability, which means:
\begin{equation}
    \mathop {\min }\limits_{{p_i'}} KL(\mathcal{P} \parallel \mathcal{Q}),
\end{equation}
subject to:
\begin{equation}
    \sum_{i=1}^C p_i' =1,
    \label{prob-constrain}
\end{equation}
\begin{equation}
    p_i' \geq 0 ,\forall i \in \{1,2,...,C\},
    \label{non-neg}
\end{equation}
\begin{equation}
    p_c' = \frac{1}{B} \sum_{i=1}^B q_c^i.
\end{equation}
We formulate the Lagrangian function $L_\lambda$ with multiplier $\lambda$ to handle the relative entropy between $\mathcal{P}$ and $\mathcal{Q}$ and related constraints, that is:
\begin{equation}
    L_\lambda(p_1', \dots, p_C', \lambda) = \sum_{i=1}^C p_i \log \frac{p_i}{p_i'} + \lambda \left(\sum_{i=1}^C p_i' - 1\right).
    \label{lagrang}
\end{equation}
When $i\neq c$, we take the partial derivative with respect to $p_i'$ in Eq. (\ref{lagrang}), that is:
\begin{equation}
    \frac{\partial L_\lambda}{\partial p_i'} = \frac{\partial}{\partial p_i'} \left( p_i \log \frac{p_i}{p_i'} \right) + \frac{\partial}{\partial p_i'} \left( \lambda \sum_{j=1}^C p_j' - \lambda \right).
\end{equation}
Compute the partial derivative of the first part, we have:
\begin{equation}
    \frac{\partial}{\partial p_i'} \left( p_i \log \frac{p_i}{p_i'} \right) = p_i \frac{\partial}{\partial p_i'} \left( \log p_i - \log p_i' \right) = -\frac{p_i}{p_i'}.
    \label{partial-first}
\end{equation}
Compute the partial derivative of the second part, we have:
\begin{equation}
    \frac{\partial}{\partial p_i'} \left( \lambda \sum_{j=1}^C p_j' - \lambda \right) = \lambda.
    \label{partial-second}
\end{equation}
Combining the two parts mentioned in Eq. (\ref{partial-first}) and Eq. (\ref{partial-second}), we have:
\begin{equation}
    \frac{\partial L_\lambda}{\partial p_i'} = -\frac{p_i}{p_i'} + \lambda.
\end{equation}
Setting the combined partial derivative to zero, we obtain:
\begin{equation}
    \frac{\partial L_\lambda}{\partial p_i'} = -\frac{p_i}{p_i'} + \lambda = 0.
\end{equation}
Hence,
\begin{equation}
    p_i' = \frac{p_i}{\lambda} \quad  (i\neq c).
    \label{pi'}
\end{equation}
Based on the derived result in Eq. (\ref{pi'}), we leverage the known $p_c'$ in Eq. (\ref{pc'}) and the probability sum constraint in Eq. (\ref{prob-constrain}) to determine $\lambda$ as follows:
\begin{equation}
    p_c' + \sum_{\substack{i = 1, i \neq c}}^{C} \frac{p_i}{\lambda} = 1.
\end{equation}
Thus,
\begin{equation}
    \lambda = \frac{\sum_{\substack{i = 1, i \neq c}}^{C} p_i}{1 - p_c'}.
\end{equation}
We substitute the value of $\lambda$ in Eq. (\ref{pi'}) to compute $p_i'$ when $i \neq c$, that is:
\begin{equation}
    p_i' = \frac{p_i}{\lambda} = \frac{p_i (1 - p_c')}{\sum_{\substack{i = 1, i \neq c}}^{C} p_i}=\frac{p_i (1 - \frac{1}{B} \sum_{i=1}^B q_c^i)}{\sum_{\substack{i = 1, i \neq c}}^{C} p_i}.
\end{equation}
Obviously, the derived results of $p_i'$ ($i \neq c$) satisfy the non-negativity constraint in Eq. (\ref{non-neg}).
\textcolor{black}{At this point, we can conclude that}:
\begin{equation}
    p_i' = \left\{ {\begin{array}{*{20}{l}}
{\frac{{{p_i}(1 - \frac{1}{B}\sum\limits_{i = 1}^B {q_c^i} )}}{{\sum_{\substack{i = 1, i \neq c}}^{C} {{p_i}} }}},&{{\rm{if }}\quad i \ne c}\\
{\frac{1}{B}\sum\limits_{i = 1}^B {q_c^i} },&{{\rm{if }}\quad i = c}
\end{array}} \right.
\label{Q-compute}
\end{equation}
\textcolor{black}{This entire process ensures that the rectified knowledge $\mathcal{Q}={[p_1',p_2',…,p_C']}$ transferred by $v^*$ is both accurate and aligned with historical predictions.}

\begin{algorithm}[t]
\newcommand{\removelatexerror}{\let\@latex@error\@gobble}
	\caption{Bridge Sample Based Online Distillation Protocol (BSBODP) with Self-Knowledge Rectification (SKR)}
	\SetAlgoNoLine

	\label{alg-bsbodp-akr}
	\begin{spacing}{0.65}
		\KwIn{$v^1$,$v^2$}
		\KwOut{Trained $Model(v^1)$, $Model(v^2)$}
        \begin{algorithmic}[0]
        
            \Procedure{BSBODP-SKR}{$v^1$, $v^2$}
           \State \Call{BSBODP-SKR-Directional}{$v^1$, $v^2$}
          \State \Call{BSBODP-SKR-Directional}{$v^2$, $v^1$}
          \State \Return{$Model(v^1)$, $Model(v^2)$}
		\EndProcedure

           \Procedure{BSBODP-SKR-Directional}{$v^{S}$, $v^{T}$}
            \State 
                $v^{T}$ generates bridge samples $dec(\varepsilon)$ from all stored embeddings $\varepsilon$
            \State $v^{T}$ extracts logits ${z^\varepsilon} = f(dec(\varepsilon );{W^{{v^T}}})$ on bridge samples
            \State $v^{T}$ computes softmax probabilities $\mathcal{P}=\tau(\frac{{{z^\varepsilon }}}{T})$
            \State \If{${p_{{y^*}}} < \tau {(\frac{{{z^\varepsilon }}}{T})_i},\exists i \in \{ 1,2,...,C\}$}
            {
                \State \If{$|\mathcal{B}_{{y^*}}|=0$}
                {
                    \State Transmits $\mathcal{P}$ to $v^{S}$
                }
                \State \Else
                {
                    \State Compute $\mathcal{Q}$ according to Eq. (\ref{Q-compute})
                    \State Transmits $\mathcal{Q}$ to $v^{S}$
                }
            }
            \State \Else
            {
                \State \If{$|\mathcal{B}_{{y^*}}|=B$}
                {
                    $\mathcal{B}_{{y^*}}$.pop()
                }
                \State $\mathcal{B}_{{y^*}}$.push($p_{{y^*}}$)
                \State Transmits $\mathcal{P}$ to $v^{S}$
            }
            \State \If{Receive $\mathcal{P}$ from $v^{T}$}
            {            
            \State \If{$v^{S} \notin \mathcal{L}$}
            {
            \State Optimize $W^{v^{S}}$ according to Eq. (\ref{optim-non-leaf})
            }
            \State \Else
            {
            \State Optimize $W^{v^{S}}$ according to Eq. (\ref{optim-leaf})
            }
            }
            \State \Else
            {
            \State \If{$v^{S} \notin \mathcal{L}$}
            {
            \State Optimize $W^{v^{S}}$ according to Eq. (\ref{new-loss-non-leaf})
            }
            \State \Else
            {
            \State Optimize $W^{v^{S}}$ according to Eq. (\ref{new-loss-leaf})
            }
            }
            \EndProcedure   
           \end{algorithmic}
	\end{spacing}
\end{algorithm}

\subsubsection{Application of SKR in BSBODP}
\textcolor{black}{To apply SKR in BSBODP, we consider a scenario where $v^*$ serves as the teacher computing node $v^T$, with its corresponding neighboring node acts as the student computing node $v^S$. For any node pairs $v^T$ and $v^S$, the rectified knowledge $\mathcal{Q}$ generated by $v^T$ is used to guide the training of $v^S$. In this case, the optimization objective for the non-leaf $v^S$ can be reformulated as:}
\begin{equation}
    \begin{array}{l}{\;\;\;\;\mathop {\min }\limits_{{W^{{v^S}}}} {L'_{non - leaf}}}\\{ = \mathop {\min }\limits_{{W^{{v^S}}}} [{L_{CE}}(\tau (f(dec(\varepsilon );{W^{{v^S}}})),{y^\varepsilon }) + }\\{\;\;\;\;\beta  \cdot KL(\tau (f(dec(\varepsilon );{W^{{v^S}}}))||{\mathcal{Q}})]}{}.\end{array}
    \label{new-loss-non-leaf}
\end{equation}
Correspondingly, the optimization objective of leaf $v^S$ is re-represented as:
\begin{equation}
    \begin{array}{l}{\;\;\;\;\mathop {\min }\limits_{{W^{{v^S}}}} {L'_{leaf}}}\\{ = \mathop {\min }\limits_{{W^{{v^S}}}} {L_{CE}}(f({X^*};{W^{{v^S}}}),{y^*}) + \gamma  \cdot {L'_{non - leaf}}.}\end{array}
    \label{new-loss-leaf}
\end{equation}
The execution procedure of BSBODP combined with SKR is listed in Algorithm \ref{alg-bsbodp-akr}. Therefore, the transferred knowledge can be adaptively rectified within each computing node before it is propagated to other nodes, improving the robustness of the entire EECC-empowered FL system.
\color{black}

\begin{algorithm}[!t]
	\caption{End-Edge-Cloud Federated Learning with Self-Rectified Knowledge Agglomeration (FedEEC)}
	\SetAlgoNoLine
	\LinesNotNumbered
	\label{alg2}
	\begin{spacing}{0.65}
		\KwIn{$\mathcal{V},\mathcal{E}$}
		\KwOut{Trained $Model(v^*), \forall v^*\in \mathcal{V}$}
  \begin{algorithmic}[0]
      \Procedure{FedEEC}{$\mathcal{V},\mathcal{E}$}
      \State \Call{Init}{$r,\mathcal{E}$}
      \State \While{do not reach maximum communication rounds}
      {
      \State \Call{FedEECTrain}{$r,\mathcal{E}$}
      }
      \EndProcedure

      \Procedure{Init}{$v^*,\mathcal{E}$}
      \State \If{$v^*=r$}
      {
      \State \For{$u \in Child(v^*)$ in parallel}{
      \State \Call{Init}{$u,\mathcal{E}$}
      \State Receive and store embeddings $\varepsilon$ with corresponding labels from $u$}
      }
      \State \ElseIf{$v^* \in \mathcal{L}$}
      {
      \State Extract and store embeddings $\varepsilon$ via $enc(X^*),\forall X^*\in \mathcal{D}^{v^*}$
      
      \State Transmit embeddings $\varepsilon$ with corresponding labels to $Parent(v^*)$
      }
      \State \Else
      {
          \For{$u \in Child(v^*)$ in parallel}
          {
          \State \Call{Init}{$u,\mathcal{E}$}
          \State Receive and store embeddings $\varepsilon$ from $u$
          \State Send embeddings $\varepsilon$ with corresponding labels to $Parent(v^*)$
          }
      }
      \EndProcedure

      \Procedure{FedEECTrain}{$v^*,\mathcal{E}$}
      \State \If{$v^*=r$}
      {
      \State \For{$u \in Child(v^*)$ in parallel}
      {
      \State \Call{FedEECTrain}{$u,\mathcal{E}$}
      }
      }
      \State \ElseIf{$v^*\in \mathcal{L}$}
      {\Call{BSBODP-SKR}{$v^*$,$Parent(v^*)$}}
      \State \Else
      {
      \State \For{$u \in Child(v^*)$ in parallel}
      {
            \State \Call{FedEECTrain}{$u,\mathcal{E}$}
            \State \Call{BSBODP-SKR}{$v^*$,$Parent(v^*)$}
      }
      }
      \EndProcedure
  \end{algorithmic}	
	\end{spacing}
\end{algorithm}

\subsection{Recursive Agglomeration in End-Edge-Cloud Networks}
\textcolor{black}{Considering FL in an EEC-NET where computing nodes are organized in a tree topology, we propose the FedEEC framework. Through the combination of BSBODP and SKR, FedEEC enables collaborative training over every pair of parent-child computing nodes,} recursively distilling knowledge from bottom to top in an agglomerative manner, as formulated in Algorithm \ref{alg2}. Specifically, FedEEC includes two main phases: initialization and recursive training with BSBODP and SKR. 
\textcolor{black}{In the initialization phase, leaf nodes generate embeddings using their pre-trained encoders and send the embeddings up the tree hierarchy to the root node.} In the training phase, each node recursively distills the knowledge from the sub-tree that seeks it as the root and passes the newly extracted knowledge to its parent, applying BSBODP combined with SKR to enable interaction between parent-child pairs.
This process starts from the leaf nodes and ends with the cloud server, where the model on the cloud is updated with the learned knowledge from all tiers. The hierarchical structure of EEC-NET, combined with the model-agnostic nature of BSBODP and SKR, allows upper-tier nodes to deploy larger models and integrate more knowledge from lower-tier nodes to achieve superior performance, and the largest model (i.e. the model on the cloud) is eventually the model that we want to gain through collaborative training.


\subsection{Deployment Flexibility Guarantees in End-Edge-Cloud Network}
FedEEC supports dynamic migration of computing nodes to guarantee deployment flexibility in an EEC-NET, where various factors such as mobility or unstable connections may require computing nodes to switch to a different parent. 
To demonstrate the advantage of FedEEC in handling dynamic migration of computing nodes, we classify existing FL interaction protocols that potentially support hierarchical collaborative model training into two types based on their constraints on model structures of the interacting computing nodes: equivalence interaction protocols and partial order interaction protocols. 
In the following part, we will give formal definitions of these two types of interaction protocols, and prove that equivalence interaction protocols, including our proposed BSBODP combined with SKR, provide better support for dynamic migration of computing nodes.

\noindent
\textbf{Definition 1.} \textit{(Equivalence Interaction Protocol). Consider an EEC-NET with a tree topology formulated in section \ref{eecc}, we define a binary relation $R$ over any pair of parent-child computing nodes.
The sufficient necessary condition for an equivalence interaction protocol is defined as follows:}
\begin{equation}
    < Model({v^i}),Model({v^i}) >  \in R,\forall {v^i} \in \mathcal{V},
\end{equation}
\begin{equation}
    \begin{array}{l}
 < Model({v^i}),Model({v^j}) >  \in R \wedge {v^i} \ne {v^j}\\
 \to  < Model({v^j}),Model({v^i}) >  \in R,\forall {v^i},{v^j} \in \mathcal{V},
\end{array}
\end{equation}
\begin{equation}
\begin{array}{l}
 < Model({v^i}),Model({v^j}) >  \in R\\
 \wedge  < Model({v^j}),Model({v^k}) >  \in R\\
 \to  < Model({v^i}),Model({v^k}) >  \in R,\forall {v^i},{v^j},{v^k} \in \mathcal{V},
\end{array}
\end{equation}
\textit{that is:}
\begin{equation}
   Model(v^i) \sim Model(v^j),\forall <v^i,v^j>\in \mathcal{E}.
\end{equation}
According to our definition, we can easily prove that the followings are equivalent interaction protocols: 1) the same model structure is adopted on the parent and child computing nodes, i.e. $Model(v^i) = Model(v^j),\forall <v^i,v^j>\in \mathcal{E}$, which is represented by FedAvg \cite{mcmahan2017communication}; 
2) model-agnostic interaction protocols that do not impose restrictions on the model structures of parent and child computing nodes, i.e. $Model(v^i) \bot Model(v^j),\forall <v^i,v^j>\in \mathcal{E}$, which is represented by BSBODP combined with SKR.

\noindent
\textbf{Definition 2.} \textit{(Partial Order Interaction Protocol). 
Consider an EEC-NET with a tree topology formulated in section \ref{eecc}, we define a binary relation $R$ over any pair of parent-child computing nodes.
The sufficient necessary condition of a partial order interaction protocol is defined as follows:}
\begin{equation}
    <Model(v^i),Model(v^i)> \in R,\forall{v^i} \in \mathcal{V},
\end{equation}
\begin{equation}
\begin{array}{l}
 < Model({v^i}),Model({v^j}) >  \in R\\
 \wedge  < Model({v^j}),Model({v^i}) >  \in R\\
 \to {v^i} = {v^j},\forall {v^i},{v^j} \in \mathcal{V},
\end{array}
\end{equation}
\begin{equation}
    \begin{array}{l}
 < Model({v^i}),Model({v^j}) >  \in R\\
 \wedge  < Model({v^j}),Model({v^k}) >  \in R\\
 \to  < Model({v^i}),Model({v^k}) >  \in R,\forall {v^i},{v^j},{v^k} \in \mathcal{V},
\end{array}
\end{equation}
\textit{that is:}
\begin{equation}
   Model(v^i) \preceq Model(v^j),\forall <v^i,v^j>\in \mathcal{E}.
\end{equation}
We can also prove that partial training-based interaction protocols \cite{alam2022fedrolex,nguyen2023feddct}, which require the model on the child node to be a sub-model of that on the parent node, i.e. $Model({v^i}) \subseteq Model(Parent({v^i})),\forall  < {v^i},Parent({v^i}) >  \in \mathcal{E}$ are partial order interaction protocols.

\noindent
\textbf{Theorem 1.} \textit{HFL methods based on equivalence interaction protocols allow the parent of any $v^1$ switch to $Parent(v^2)$ while preserving both operational integrity and interaction consistency between computing nodes in the EEC-NET, where $v^1$ and $v^2$ are non-root nodes, i.e. $Model({v^1}) \sim Model(Parent({v^2})),\forall {v^1},{v^2} \in \mathcal{V}- \mathcal{V}_1$.}

\noindent
\textit{Proof.} \\
\noindent 
First, we prove that any two computing nodes in the EEC-NET satisfy the equivalence relation determined by $R$. Without loss of generality, we have two arbitrary non-root nodes $v^x,v^y \in \mathcal{V} - \mathcal{V}_1$ in the EEC-NET. Since $\mathcal{V}$ is a tree topology, any two nodes are connected. Therefore, there must exist a finite sequence of $m$ adjacent intermediate nodes $v^*_1,v^*_2,...,v^*_m$ such that $v^x$ and $v^y$ are connected through them, that is:
\begin{equation}
    Model(v^x)\sim Model(v^*_1),
\end{equation}
\begin{equation}
    Model(v^y)\sim Model(v^*_m),
\end{equation}
\begin{equation}
    Model(v^*_{i-1})\sim Model(v^*_i),\forall i \in \{2,3,...,m\}.
\end{equation}
Due to the reflexivity and transitivity properties of equivalence relations, we can infer that:
\begin{equation}
    \begin{array}{l}
         Model(v^x)\sim Model(v^*_1)\sim Model(v^*_2) \\ \sim ... \sim Model(v^*_m) \sim Model(v^y).
    \end{array}
\end{equation}
Thus, we can conclude that:
\begin{equation}
    Model(v^x)\sim Model(v^y).
\end{equation}

\noindent
Next, we prove that $v^1$ can change its parent to $Parent(v^2)$ while maintaining network integrity. Since both $v^1$ and $Parent(v^2)$ are distributed in $\mathcal{V}$, they satisfy an equivalence relation according to the aforementioned proof. Hence,
\begin{equation}
    Model(v^1) \sim Model(Parent(v^2)).
\end{equation}
Therefore, the computing node $v^1$ is allowed to become a child of $Parent(v^2)$, i.e. dynamic migration of computing nodes is allowed. 
$\hfill \square$



\noindent
\textbf{Theorem 2.} \textit{HFL methods based on partial order interaction protocols do not necessarily allow the parent of any $v^1$ switch to $Parent(v^2)$ while preserving both operational integrity and interaction consistency between computing nodes in the EEC-NET, where $v^1$ and $v^2$ are non-root nodes, i.e. $\neg Model({v^1}) \preceq Model(Parent({v^2})),\exists {v^1},{v^2} \in \mathcal{V}- \mathcal{V}_1$}.

\noindent
\textit{Proof.} \\
We demonstrate that when $v^1$ and $v^2$ are at the same tier, partial order interaction protocols already no longer necessarily allow parent switching for $v^1$ while maintaining satisfaction.\\
\textbf{Case 2.1.} When ${v^1},{v^2} \in {\mathcal{V}_2}$, we have:
\begin{equation}
    Parent(v^1) = Parent(v^2) = r.
\end{equation}
Hence, $v^1$ and $v^2$ have the same parent node $r$, and there is no dynamic migration of computing nodes in this case.

\noindent
\textbf{Case 2.2.} When ${v^1},{v^2} \in {\mathcal{V}_t},t \ge  3$, there are two sub-cases:
\begin{equation}
    Model(Parent(v^1)) \preceq Model(Parent(v^2)),
\end{equation}
and
\begin{equation}
    Model(Parent(v^2)) \preceq Model(Parent(v^1)).
    \label{case2}
\end{equation}
When Eq. (\ref{case2}) is satisfied, there exists a situation where computing node $v^1$ is not allowed to switch its parent to $Parent(v^2)$.
Instantiating $\preceq$ to the partial order relation over integers $\le$, we construct a tree topology $10(9(8,7),5(4,3))$ and set function $Model(\cdot)$ to be a constant function, i.e. $Model(x)=x,\forall x$. 
In our setting, the parent of 7 ($v^1$) and 3 ($v^2$) are at the same tier and $Model(Parent(3))=5\le9=Model(Parent(7))$ satisfies Eq. (\ref{case2}). At this point, $\neg Model(7)\le Model(Parent(3))$ is equivalent to $\neg 7\le 5$, which is apparently true. Hence, dynamic migration of computing nodes is not necessarily allowed.
$\hfill \square$

\section{Experiments}
\subsection{Experiment Setup}
\subsubsection{Hardware and Software Platforms}
We conduct experiments on a Huawei 2288H V5 physical server equipped with two Intel(R) Xeon(R) Gold 5218 CPUs, two NVIDIA RTX 3090 GPUs, and 384 GB AM. To comprehensively evaluate the performance of our proposed FedEEC, we adopt the FedML research library \cite{he2020fedml} deployed on Linux 6.8.0-38-generic, as the software platform for all experiments.

\subsubsection{Basic Settings} 
We select three representative datasets, SVHN \cite{netzer2011reading}, CIFAR-10 \cite{cifar10}, and CINIC-10 \cite{cinic10} to assess the performance of FedEEC under varying data scenarios. The primary evaluation metric used in this paper is cloud model accuracy. Regarding model structures, we deploy four types of models for collaborative training, which are $M_{end}^1$, $M_{end}^2$, $M_{edge}$, and $M_{cloud}$. Specifically, $M_{end}^1$ and $M_{end}^2$ are lightweight, three-layer Convolutional Neural Networks (CNNs) differing in the size of their intermediate layers. 
$M_{edge}$ and $M_{cloud}$ are Deep Residual Networks (ResNet) \cite{he2016deep}, designed with varying depths to suit the computational capabilities of edge and cloud nodes.
Additionally, we introduce a six-layer autoencoder, $M_{auto}$ for auxiliary training. $M_{auto}$ comprises  a three-layer encoder ($M_{enc}$) and a three-layer decoder ($M_{dec}$), pre-trained on ImageNet \cite{deng2009imagenet} dataset. The main configurations of adopted models are provided in Table \ref{model-config}.
To ensure fair comparisons, we control the key experimental configurations to isolate the impact of the HFL algorithms. The number of devices, data distributions, learning rate, and batch size are kept consistent across the same group of experiments to minimize the influence of non-algorithmic factors.

\subsubsection{Baseline Algorithms}
We compare FedEEC against the following five HFL or distributed learning algorithms:
\begin{itemize}
    \item HierFAVG \cite{liu2020client}, which is a classical client-edge-cloud FL framework that introduces edge servers for partial model aggregation, reducing communication overhead and shortening training time.
    \item DemLearn \cite{nguyen2023self}, which is a distributed learning method for complex learning tasks with adaptive hierarchical organization.
    \item HierMo \cite{yang2023hierarchical}, which is an HFL method incorporating momentum acceleration, which accelerates model convergence by aggregating momentum across clients and edge servers.
    \item HierQSGD \cite{liu2022hierarchical}, which is an HFL method that integrates model quantization techniques to reduce communication overhead.
    \item FedAgg \cite{wu2024agglomerative}, which is the previous generation version of FedEEC, allowing trained models from end, edge to cloud nodes to grow larger in size and complexity through model-agnostic multi-tier collaboration.
\end{itemize}

\subsection{Implementation Details}
\subsubsection{Simulating Non-IID Data Distribution}
To simulate real-world non-independent and identically distributed (non-IID) on-device data, we utilize the data partition module in FedML to create heterogeneous data distributions. \textcolor{black}{The degree of data heterogeneity is controlled by the hyperparameter $\alpha=2.0$, which corresponds to the concentration parameter of the Dirichlet distribution used for partitioning data among devices \cite{li2022federated}.} For each experiment, we record the cloud model accuracy on the test dataset over 100 communication rounds, using the maximum accuracy observed during training as the final performance metric.

\subsubsection{Scaling Across Varying Number of Devices}
To evaluate the scalability of FedEEC with varying numbers of devices, we conduct experiments over 50, 100, and 500 devices. For each configuration, we set up 5, 10, and 20 edge servers, respectively, with devices evenly distributed across these servers. In the exceptional case of DemLearn, the number of edge servers is determined by the algorithm’s adaptive organization process.

\subsubsection{Hierarchical Model Deployment}
For multi-tier aggregation algorithms (HierFAVG, DemLearn, HierMo, and HierQSGD), identical model structures are required across all computing nodes. Given the limited computational power of end devices, we uniformly deploy $M_{end}^1$ on end, edge, and cloud nodes for these algorithms. In contrast, for FedAgg and FedEEC that do not require model aggregation, we deploy $M_{end}^1$, $M_{edge}$, and $M_{cloud}$ at end, edge, and cloud nodes, respectively, to accommodate different computational capabilities across tiers. Additionally, we deploy the encoder ($M_{enc}$) on the end devices and distribute the decoder ($M_{dec}$) across all nodes to assist in the training process of FedAgg and FedEEC.

\subsubsection{Baseline Comparisons}
We conduct comprehensive evaluations of FedEEC against all considered baseline algorithms over SVHN and CIFAR-10 datasets. For CINIC-10 dataset, we compare FedEEC with HierFAVG, HierMo, HierQSGD, and FedAgg. Due to the significant computational overhead introduced by the adaptive organization process in DemLearn, it becomes impractical to apply this method to such a large-scale dataset.

\subsubsection{Hyperparameters}
In all experiments, we set the learning rate to 0.001 and the batch size to 8. The specific hyperparameter configurations for each algorithm are as follows:
\begin{itemize}
    \item 
    For HierFAVG, we set $\kappa_1=1$ and $\kappa_2=1$.
    \item 
    For DemLearn, we set $K\_{Levels}=2$ to form a three-tier end-edge-cloud architecture. Other hyperparameters follow the default setting in \cite{demlearn-code}. 
    \item 
    For HierQSGD, we set $\kappa_1=1$, $\kappa_2=1$, $q\_de=1.0$, $q\_ec=1.0$. Other hyperparameters follow the default setting in \cite{hierqsgd-code}.
    \item 
    For HierMo, we set $\tau=1$, $\pi=1$, $\gamma=0.5$ and $\gamma_a=0.9$.
    \item 
    For FedAgg, we set $\gamma=1.0$, $T=0.5$, and $\beta=1.5$.
    \item 
    For FedEEC, we set $B=20$, $\gamma=1.0$, $T=0.5$, and $\beta=1.5$.
\end{itemize}

\begin{table}[]
\centering
\caption{Model configurations used in experiments.}
\renewcommand{\arraystretch}{0.9}
\label{model-config}
\begin{tabular}{l|c|c|c}
\hline
\multicolumn{1}{c|}{\textbf{\begin{tabular}[c]{@{}c@{}}Model\end{tabular}}} & \multicolumn{1}{l|}{\textbf{Notation}} & \textbf{\begin{tabular}[c]{@{}c@{}}Deployment\end{tabular}} & \textbf{\begin{tabular}[c]{@{}c@{}}Parameters\end{tabular}} \\ \hline
CNN-1                                                                              & $M_{end}^1$                                  & End, Edge, Cloud                                                       & 12.84K                                                               \\
CNN-2                                                                              & $M_{end}^2$                                  & End, Edge, Cloud                                                       & 11.67K                                                               \\
ResNet-10                                                                          & $M_{edge}$                                  & Edge, Cloud                                                            & 4.68M                                                                \\
ResNet-18                                                                          & $M_{cloud}$                                 & Cloud                                                                  & 10.66M                                                               \\
Encoder                                                                            & $M_{enc}$                                   & End                                                                    & 1.90K                                                                \\
Decoder                                                                            & $M_{dec}$                                   & End, Edge, Cloud                                                       & 2.47K                                                                \\ \hline
\end{tabular}
\end{table}

\begin{table*}[]
\centering
\caption{Comparison of cloud model accuracy (\%) across different datasets and client numbers. \textbf{Bold} values represent the best accuracy, the same as below.}
\label{main-exp}
\setlength{\tabcolsep}{8pt}
\renewcommand{\arraystretch}{0.9}
\begin{tabular}{c|l|ccc|ccc|c}
\hline \hline
\multirow{8}{*}{\textbf{SVHN}}     & \multicolumn{1}{c|}{\multirow{2}{*}{\textbf{Method}}} & \multicolumn{3}{c|}{\textbf{Trained Model}}                               & \multirow{2}{*}{\textbf{50 Clients}} & \multirow{2}{*}{\textbf{100 Clients}} & \multirow{2}{*}{\textbf{500 Clients}} & \multirow{2}{*}{\textbf{Average}} \\
                                   & \multicolumn{1}{c|}{}                                 & \textbf{End}           & \textbf{Edge}          & \textbf{Cloud}          &                                      &                                       &                                       &                                   \\ \cline{2-9} 
                                   & HierFAVG                                              & \multicolumn{3}{c|}{\multirow{4}{*}{$M_{end}^1$}}                               & 19.59                                & 28.85                                 & 19.91                                 & 22.78                             \\
                                   & DemLearn                                              & \multicolumn{3}{c|}{}                                                     & 30.75                                & 28.31                                 & 22.25                                 & 27.10                             \\
                                   & HierMo                                                & \multicolumn{3}{c|}{}                                                     & 19.55                                & 19.59                                 & 20.30                                 & 19.81                             \\
                                   & HierQSGD                                              & \multicolumn{3}{c|}{}                                                     & 19.59                                & 19.59                                 & 19.59                                 & 19.59                             \\ \cline{2-9} 
                                   & FedAgg                                                & \multirow{2}{*}{$M_{end}^1$} & \multirow{2}{*}{$M_{edge}$} & \multirow{2}{*}{$M_{cloud}$} & 72.05                                & 78.54                                 & 79.67                                 & 76.75                             \\
                                   & \textbf{FedEEC}                                       &                        &                        &                         & \textbf{80.33}                       & \textbf{79.98}                        & \textbf{81.85}                        & \textbf{80.72}                    \\ \hline \hline
\multirow{8}{*}{\textbf{CIFAR-10}} & \multicolumn{1}{c|}{\multirow{2}{*}{\textbf{Method}}} & \multicolumn{3}{c|}{\textbf{Trained Model}}                               & \multirow{2}{*}{\textbf{50 Clients}} & \multirow{2}{*}{\textbf{100 Clients}} & \multirow{2}{*}{\textbf{500 Clients}} & \multirow{2}{*}{\textbf{Average}} \\
                                   & \multicolumn{1}{c|}{}                                 & \textbf{End}           & \textbf{Edge}          & \textbf{Cloud}          &                                      &                                       &                                       &                                   \\ \cline{2-9} 
                                   & HierFAVG                                              & \multicolumn{3}{c|}{\multirow{4}{*}{$M_{end}^1$}}                               & 19.23                                & 14.64                                 & 19.14                                 & 17.67                             \\
                                   & DemLearn                                              & \multicolumn{3}{c|}{}                                                     & 30.71                                & 27.66                                 & 20.64                                 & 26.34                             \\
                                   & HierMo                                                & \multicolumn{3}{c|}{}                                                     & 16.94                                & 10.78                                 & 16.12                                 & 14.61                             \\
                                   & HierQSGD                                              & \multicolumn{3}{c|}{}                                                     & 13.85                                & 13.22                                 & 10.00                                 & 12.36                             \\ \cline{2-9} 
                                   & FedAgg                                                & \multirow{2}{*}{$M_{end}^1$} & \multirow{2}{*}{$M_{edge}$} & \multirow{2}{*}{$M_{cloud}$} & 31.67                                & 33.66                                 & 34.61                                 & 33.31                             \\
                                   & \textbf{FedEEC}                                       &                        &                        &                         & \textbf{33.46}                       & \textbf{33.96}                        & \textbf{35.42}                        & \textbf{34.28}                    \\ \hline \hline
\multirow{7}{*}{\textbf{CINIC-10}} & \multicolumn{1}{c|}{\multirow{2}{*}{\textbf{Method}}} & \multicolumn{3}{c|}{\textbf{Trained Model}}                               & \multirow{2}{*}{\textbf{50 Clients}} & \multirow{2}{*}{\textbf{100 Clients}} & \multirow{2}{*}{\textbf{500 Clients}} & \multirow{2}{*}{\textbf{Average}} \\
                                   & \multicolumn{1}{c|}{}                                 & \textbf{End}           & \textbf{Edge}          & \textbf{Cloud}          &                                      &                                       &                                       &                                   \\ \cline{2-9} 
                                   & HierFAVG                                              & \multicolumn{3}{c|}{\multirow{3}{*}{$M_{end}^1$}}                               & 12.32                                & 11.40                                 & 10.91                                 & 11.54                             \\
                                   & HierMo                                                & \multicolumn{3}{c|}{}                                                     & 12.08                                & 11.01                                 & 16.24                                 & 13.11                             \\
                                   & HierQSGD                                              & \multicolumn{3}{c|}{}                                                     & 13.65                                & 16.07                                 & 12.09                                 & 13.94                             \\ \cline{2-9} 
                                   & FedAgg                                                & \multirow{2}{*}{$M_{end}^1$} & \multirow{2}{*}{$M_{edge}$} & \multirow{2}{*}{$M_{cloud}$} & 15.24                                & 16.58                                 & 17.99                                 & 16.60                             \\
                                   & \textbf{FedEEC}                                       &                        &                        &                         & \textbf{16.96}                       & \textbf{17.16}                        & \textbf{18.44}                        & \textbf{17.52}                    \\ \hline \hline
\end{tabular}
\end{table*}

\subsection{Evaluation Results}
\subsubsection{Performance Across Different Datasets}
Table \ref{main-exp} presents the cloud model accuracy achieved by FedEEC across various client numbers on three datasets. As shown, FedEEC consistently demonstrates strong performance across all datasets, outperforming the baseline methods in every case. On the SVHN dataset, FedEEC achieves an average accuracy of 80.72\%, significantly exceeding the performances of HierFAVG (22.78\%), DemLearn (27.10\%), HierMo (19.81\%), HierQSGD (19.59\%), and FedAgg (76.75\%). \textcolor{black}{Similarly, FedEEC achieves average accuracies of 34.8\% and 17.52\% on the more challenging CIFAR-10 and CINIC-10 datasets, significantly outperforming the best baseline methods, which achieve 33.31\% and 16.60\% respectively.} These results highlight FedEEC’s consistent state-of-the-art performance across different datasets, demonstrating its robustness and adaptability to diverse data scenarios.


\subsubsection{Performance Across Different Number of Clients}
FedEEC exhibits robust performance across varying client number settings. As shown in Table \ref{main-exp}, FedEEC consistently outperforms the baseline methods across all datasets regardless of the number of clients. Notably, within the same dataset, FedEEC maintains stable cloud model accuracy, with fluctuations less than 2\%. This robustness across varying client numbers underscores FedEEC's ability to effectively manage knowledge transfer and collaboration across tiers, even as data is distributed over a larger number of devices.
The superior performance of FedEEC across varying client numbers demonstrates its reliability and scalability as a solution for FEL.

\subsubsection{Effectiveness of Self-Knowledge Rectification}
The effectiveness of SKR mechanism is evident when comparing FedEEC (with SKR) with FedAgg (without SKR) in Table \ref{main-exp}. As demonstrated by the superior performance of FedEEC, SKR plays a crucial role in ensuring reliable knowledge transfer through the hierarchy. By mitigating the impact of misleading attributes within the knowledge, SKR improves the overall quality of the knowledge transferred, resulting in more accurate and stable model updates across tiers.

\subsubsection{Convergence Rate}
\begin{figure}[t]
	\centering
	\includegraphics[width=0.50\textwidth]{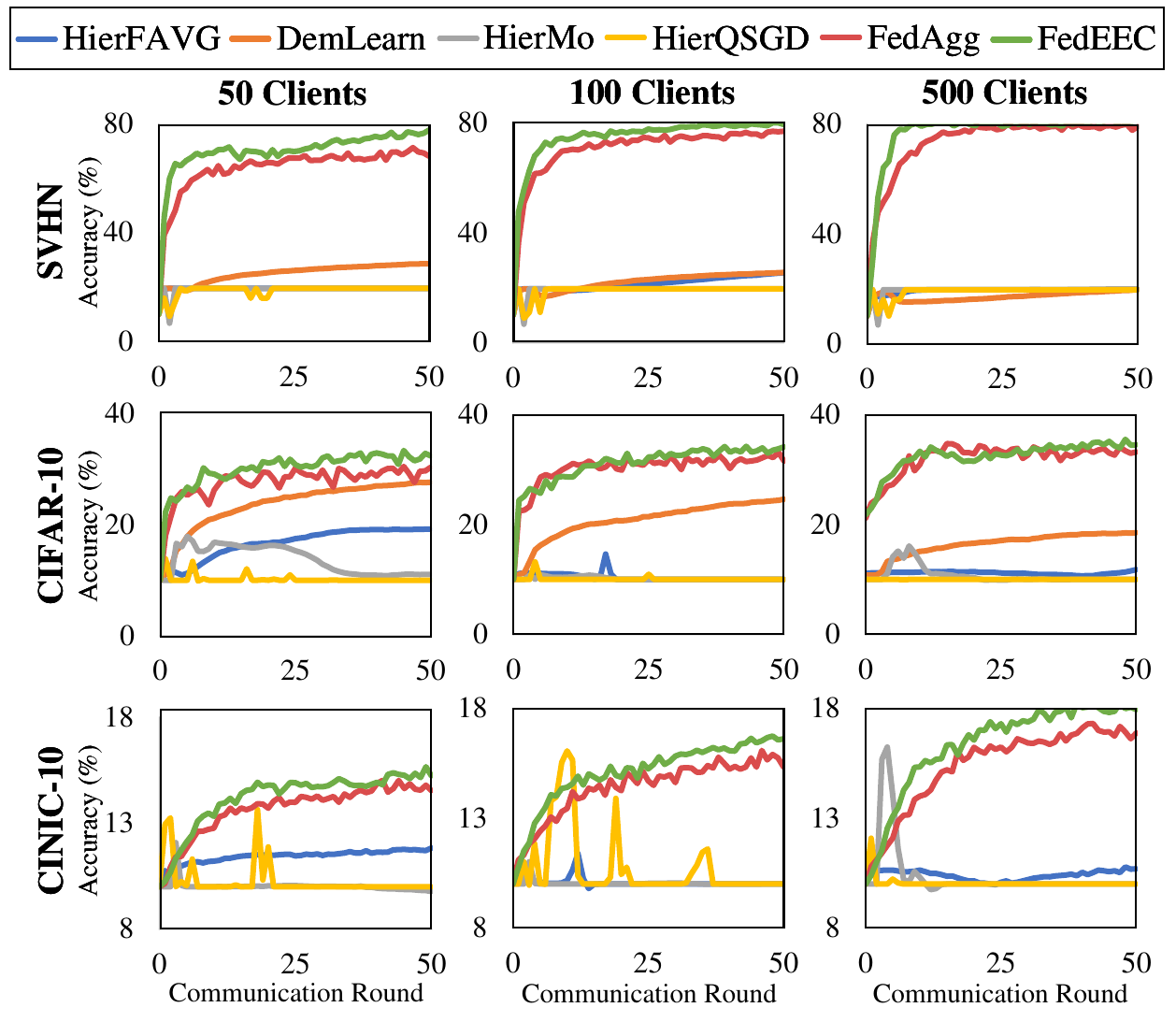}
	\caption{\textcolor{black}{Learning curves illustrating cloud model accuracy (\%) across communication rounds.}}
	\label{learning-curve}
\end{figure}

Figure \ref{learning-curve} illustrates the convergence rate of FedEEC compared to the baseline algorithms. Over the course of 50 communication rounds, FedEEC consistently converges faster than all other methods. This can be attributed to its knowledge agglomeration process, which allows the model scale growth across tiers, and the SKR mechanism, which ensures robust knowledge transfer among neighboring nodes. These features contribute to FedEEC's ability to collaboratively train larger models than those typically supported by end devices, while also reducing the propagation of suboptimal updates throughout the hierarchy.

\begin{figure}[t]
	\centering
	\includegraphics[width=0.5\textwidth]{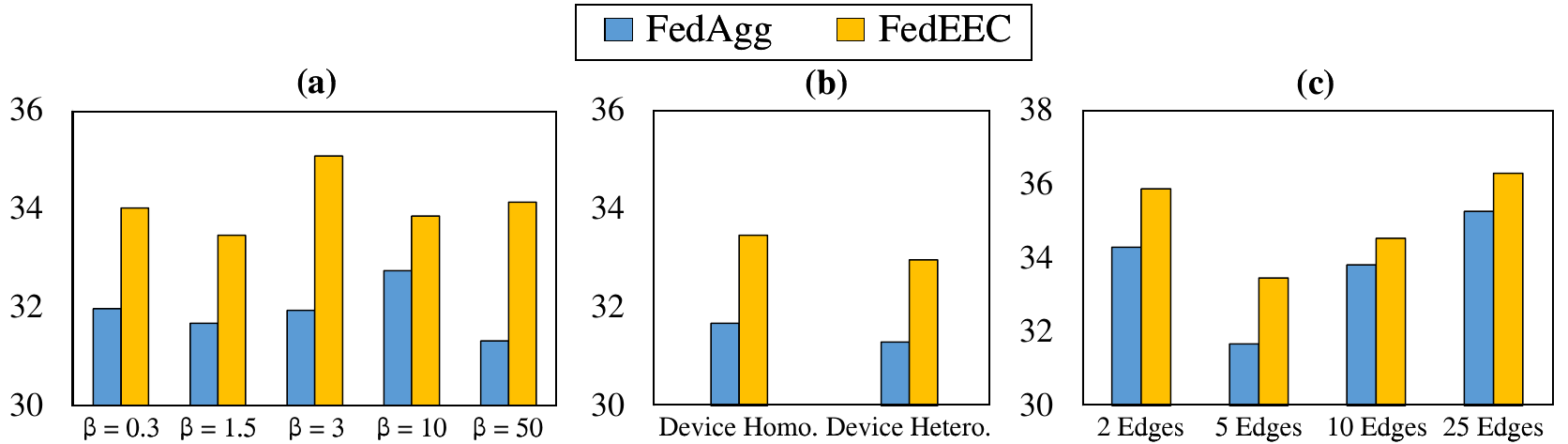}
	\caption{Cloud model accuracy (\%) of FedAgg and FedEEC over different $\beta$, on-device model setting and edge numbers.}
	\label{bar-chart}
\end{figure}

\section{Ablation Study}
In the following subsections, we use the same default hyperparameter configurations as described in the main experiments unless explicitly stated otherwise. 

\subsection{Impact of Distillation Weight}
We evaluate the effectiveness of FedEEC across various values of the distillation weight $\beta$. As shown in Figure \ref{bar-chart} (a) and Table \ref{ablation-beta}, FedEEC maintains a clear advantage over FedAgg across the entire range of $\beta$ values (from $0.3$ to $50$) with only minor performance fluctuations. This stability highlights FedEEC's robustness to variations in $\beta$, reducing the necessity for extensive hyperparameter tuning to achieve strong performance.

\subsection{Tolerance to Device Heterogeneity}
\textcolor{black}{We evaluate the robustness of FedEEC in environments with two settings of device heterogeneity.} As illustrated in Figure \ref{bar-chart} (b) and Table \ref{ablation-model}, FedEEC consistently outperforms FedAgg across both homogeneous and heterogeneous device configurations. This highlights FedEEC’s adaptability to environments with varying computational capabilities across devices, making it particularly suitable for real-world applications where seamless collaboration among differentiated on-device models is essential.

\begin{table}[]
\centering
\caption{Impact of distillation weight $\beta$ on cloud model accuracy. Results are obtained on CIFAR-10 dataset with $K=100$.}
\label{ablation-beta}
\setlength{\tabcolsep}{4pt}
\renewcommand{\arraystretch}{0.9}
\begin{tabular}{l|ccccc}
\hline
\multicolumn{1}{c|}{\textbf{Method}} & \textbf{$\beta=0.3$} & \textbf{$\beta=1.5$} & \textbf{$\beta=3$}   & \textbf{$\beta=10$}  & \textbf{$\beta=50$}  \\ \hline
FedAgg                      & 31.97          & 31.67          & 31.93          & 32.74          & 31.31          \\
\textbf{FedEEC}                      & \textbf{34.01} & \textbf{33.46} & \textbf{35.07} & \textbf{33.85} & \textbf{34.13} \\ \hline
\end{tabular}
\end{table}

\subsection{Impact of Edge Number}
We further assess the scalability of FedEEC by varying the number of edge nodes, as shown in Figure \ref{bar-chart} (c) and Table \ref{ablation-edge}. Across all edge configurations, FedEEC consistently achieves higher cloud model accuracy than FedAgg. These results demonstrate that FedEEC not only makes sufficient utilization of edge-side resources but also maintains robust performance across diverse network topologies.

\begin{table}[t]
\centering
\caption{Impact of different on-device model heterogeneity setup on cloud model accuracy. The $M^1_{end}$ and $M^2_{end}$ each constitute half of the end-side models in the device heterogeneity setup.}
\label{ablation-model}
\setlength{\tabcolsep}{9pt}
\renewcommand{\arraystretch}{0.9}
\begin{tabular}{l|cc}
\hline
\multicolumn{1}{c|}{\textbf{Method}} & \textbf{Device Homo.} & \textbf{Device Hetero.} \\ \hline
FedAgg                               & 31.67                 & 31.29                   \\
\textbf{FedEEC}                      & \textbf{33.46}        & \textbf{32.96}          \\ \hline
\end{tabular}
\end{table}

\begin{table}[t]
\centering
\caption{Impact of different edge node numbers on cloud model accuracy.}
\label{ablation-edge}
\setlength{\tabcolsep}{5pt}
\renewcommand{\arraystretch}{0.9}
\begin{tabular}{l|cccc}
\hline
\multicolumn{1}{c|}{\textbf{Method}} & \textbf{2 Edges} & \textbf{5 Edges} & \textbf{10 Edges} & \textbf{25 Edges} \\ \hline
FedAgg                               & 34.29            & 31.67            & 33.81             & 35.26             \\
\textbf{FedEEC}                      & \textbf{35.87}   & \textbf{33.46}   & \textbf{34.53}    & \textbf{36.29}    \\ \hline
\end{tabular}
\end{table}

\begin{table*}[]
\centering
\caption{Comparison of Communication Overhead Between HierFAVG and FedEEC. $r$ and $E$ represent the communication round and the number of edge servers, respectively. Both methods are evaluated under the goal of training a model with identical structure $M_{cloud}$ on the cloud.}
\label{communication-comparison}
\setlength{\tabcolsep}{5pt}
\renewcommand{\arraystretch}{0.9}
\begin{tabular}{c|l|c|ccc|c}
\hline
\hline
\multirow{6}{*}{\textbf{\begin{tabular}[c]{@{}c@{}}End-Edge\\ Communication\end{tabular}}}   & \multicolumn{1}{c|}{\textbf{Method}} & \textbf{Communication Complexity}                                                                   & \textbf{SVHN} & \textbf{CIFAR-10} & \textbf{CINIC-10} & \textbf{Average} \\ \cline{2-7} 
                                                                                             & HierFAVG                             & $O(r\cdot\sum\limits_{i = 1}^K {|{W^i}|} )$                                                             & 208.20G       & 208.20G           & 208.20G      & 208.20G     \\ \cline{2-7} 
                                                                                             & FedEEC                               & $O(\sum\limits_{k = 1}^K {|{\mathcal{D}^k}|}  \cdot (|\varepsilon |+1 + r \cdot   (|{z^\varepsilon }| + 1)))$ & \textbf{18.10G}        & \textbf{12.35G}            & \textbf{22.24G}   & \textbf{17.56G}         \\ \hline
                                                                                             \hline
\multirow{6}{*}{\textbf{\begin{tabular}[c]{@{}c@{}}Edge-Cloud\\ Communication\end{tabular}}} & \multicolumn{1}{c|}{\textbf{Method}} & \textbf{Communication Complexity}                                                                   & \textbf{SVHN} & \textbf{CIFAR-10} & \textbf{CINIC-10} & \textbf{Average} \\ \cline{2-7} 
                                                                                             & HierFAVG                             & $O(r\cdot\sum\limits_{i = 1}^E {|{W^i}|} )$                                                             & 20.82G        & 20.82G            & \textbf{20.82G}   & 20.82G         \\ \cline{2-7} 
                                                                                             & FedEEC                               & $O(\sum\limits_{k = 1}^K {|{\mathcal{D}^k}|}  \cdot (|\varepsilon |+1 + r \cdot   (|{z^\varepsilon }| + 1)))$ & \textbf{18.10G}        & \textbf{12.35G}            & 22.24G      & \textbf{17.56G}       \\ \hline
                                                                                             \hline
\end{tabular}
\end{table*}

\section{Discussions}
\subsection{Analysis on Communication Overhead}
Communication overhead is a critical factor in FL, especially in hierarchical settings like EECC. In Table \ref{communication-comparison}, we compare the network bandwidth consumption of HierFAVG and FedEEC. As shown, FedEEC demonstrates significantly better communication efficiency, reducing network bandwidth usage by 91.57\% on average in end-edge communication and 15.66\% in edge-cloud communication. This improvement is primarily due to FedEEC’s strategy of exchanging lightweight logits and embeddings instead of full model parameters, which is the standard approach in prevailing HFL methods including HierFAVG. By minimizing the size of transmitted information during training, FedEEC not only achieves robust model performance but also effectively reduces communication overhead across the network hierarchy.


\subsection{Limitations}
One limitation of FedEEC is its reliance on the pre-trained autoencoder for generating bridge samples. On the one hand, the degree to which the autoencoder aligns with the data distribution on end devices may impact the quality of the generated embeddings as well as synthetic samples, thereby affecting the overall effectiveness of the online distillation process. On the other hand, the need for large-scale pre-training introduces a dependency that may not be practical in all deployment scenarios. Another limitation of FedEEC is that the performance of these upper-tier models is fundamentally constrained by the knowledge distilled from the resource-constrained end devices. When the end devices have extremely limited computational resources, the quality of the knowledge they produce may not be sufficient to fully exploit the capacity of the larger models deployed at upper tiers. Addressing these limitations would significantly enhance FedEEC’s applicability across wider range of deployment contexts.

\section{Related Works}
\subsection{Hierarchical Federated Learning}
HFL leverages multi-tier tree topologies to manage the training process of conventional FL, coordinating clients through a central server (typically located in the cloud) and multiple intermediate computing nodes (typically located at the edge) \cite{duan2023combining}. Liu et al. \cite{liu2020client} introduced the pioneering HFL framework that facilitated faster model training and reduced edge-cloud communication costs by enabling partial model aggregation at edge servers. Subsequent enhancements include the application of quantization techniques to reduce bandwidth usage \cite{liu2023hierarchical} and the incorporation of momentum acceleration to speed up convergence \cite{yang2023hierarchical}. Research conducted by Nguyen et al. \cite{nguyen2023self} and Wang et al. \cite{wang2023accelerating} explored self-organizing structures and dynamic training topologies in HFL to improve generalization performance on heterogeneous data and reduce completion time over heterogeneous devices, respectively. Security enhancements in HFL have been addressed by Guo et al. \cite{guo2023privacy}, who implemented a two-tier differential privacy mechanism to balance efficiency and privacy.

\subsection{Training Larger Model in Federated Learning}
Prevailing FL methods \cite{mcmahan2017communication,li2020federated,karimireddy2020scaffold} typically employ homogeneous models across diverse computing nodes, which restricts the training of large-scale models due to the bottleneck effect imposed by resource-constrained devices. To overcome this limitation, techniques such as knowledge distillation and partial training have been employed to transfer insights from smaller models trained on end devices to larger models on remote servers. Partial training-based approaches \cite{alam2022fedrolex,nguyen2023feddct} divide the complete global model into multiple sub-models, training these sub-models on multiple clients in parallel. This allows the collaboratively trained server-side model to exceed the size of the largest model on any individual client. Knowledge distillation-based approaches \cite{he2020group,cho2022heterogeneous,wu2024agglomerative} use the model outputs from clients as a regularization term for server-side model training, facilitating the transfer of integrated representations learned by clients on decentralized private data.

\subsection{Knowledge Distillation Empowered Federated Learning for Edge Computing}
Knowledge distillation has become increasingly significant in addressing challenges encountered when deploying FL in edge computing scenarios \cite{wu2023survey}, such as user personalization, resource heterogeneity, and communication efficiency. Personalized or multi-task FL approaches proposed by Wu et al. \cite{wu2023fedict} and Jin et al. \cite{jin2023personalized} leverage device-specific distillation to enhance model personalization performance on edge devices. Global knowledge distillation methods proposed by Yao et al. \cite{yao2024fedgkd} mitigate client drift by using historical global models as teachers to guide local model training. Liu et al. \cite{liu2024adaptive} combined FedAvg with block-wise regularization and knowledge distillation to enhance both model performance and communication efficiency. Other notable advancements include knowledge cache-driven FL \cite{wu2024fedcache,pan2024fedcache}, which significantly improves training communication efficiency while maintaining personalization accuracy.

\section{Conclusion}
In this paper, we propose End-Edge-Cloud Federated Learning with Self-Rectified Knowledge Agglomeration (FedEEC), a novel framework that overcomes the key limitations of conventional Hierarchical Federated Learning (HFL) within End-Edge-Cloud Collaboration (EECC) environments. FedEEC effectively tackles critical challenges, including scalable model training across multi-tier computing nodes, effective hierarchical knowledge transfer, and support for dynamic node migration. By introducing the Bridge Sample-Based Online Distillation Protocol (BSBODP) and Self-Knowledge Rectification (SKR), \textcolor{black}{FedEEC facilitates model growth in size and generalization ability across tiers, while simultaneously refining the transferred knowledge to mitigate the propagation of misleading predictions and enhances cloud model optimization.} Additionally, FedEEC allows for flexible parent switching over non-root computing nodes, making the framework highly adaptive and resilient to a wide range of resource-dynamic environments. Extensive experiments across diverse datasets and configurations demonstrate that FedEEC consistently outperforms existing methods, achieving superior cloud model accuracy while maintaining communication efficiency.

\color{black}
\ifCLASSOPTIONcaptionsoff
  \newpage
\fi

\bibliographystyle{IEEEtran}

\begin{thebibliography}{10}
\providecommand{\url}[1]{#1}
\csname url@samestyle\endcsname
\providecommand{\newblock}{\relax}
\providecommand{\bibinfo}[2]{#2}
\providecommand{\BIBentrySTDinterwordspacing}{\spaceskip=0pt\relax}
\providecommand{\BIBentryALTinterwordstretchfactor}{4}
\providecommand{\BIBentryALTinterwordspacing}{\spaceskip=\fontdimen2\font plus
\BIBentryALTinterwordstretchfactor\fontdimen3\font minus
  \fontdimen4\font\relax}
\providecommand{\BIBforeignlanguage}[2]{{%
\expandafter\ifx\csname l@#1\endcsname\relax
\typeout{** WARNING: IEEEtran.bst: No hyphenation pattern has been}%
\typeout{** loaded for the language `#1'. Using the pattern for}%
\typeout{** the default language instead.}%
\else
\language=\csname l@#1\endcsname
\fi
#2}}
\providecommand{\BIBdecl}{\relax}
\BIBdecl

\bibitem{xu2021edge}
D.~Xu, T.~Li, Y.~Li, X.~Su, S.~Tarkoma, T.~Jiang, J.~Crowcroft, and P.~Hui,
  ``Edge intelligence: Empowering intelligence to the edge of network,''
  \emph{Proceedings of the IEEE}, vol. 109, no.~11, pp. 1778--1837, 2021.

\bibitem{yang2022edge}
B.~Yang, B.~Wu, Y.~You, C.~Guo, L.~Qiao, and Z.~Lv, ``Edge intelligence based
  digital twins for internet of autonomous unmanned vehicles,'' \emph{Software:
  Practice and Experience}, 2022.

\bibitem{nasir2022enabling}
M.~Nasir, K.~Muhammad, A.~Ullah, J.~Ahmad, S.~W. Baik, and M.~Sajjad,
  ``Enabling automation and edge intelligence over resource constraint iot
  devices for smart home,'' \emph{Neurocomputing}, vol. 491, pp. 494--506,
  2022.

\bibitem{gong2020edgerec}
Y.~Gong, Z.~Jiang, Y.~Feng, B.~Hu, K.~Zhao, Q.~Liu, and W.~Ou, ``Edgerec:
  recommender system on edge in mobile taobao,'' in \emph{Proceedings of the
  29th ACM International Conference on Information \& Knowledge Management},
  2020, pp. 2477--2484.

\bibitem{antunes2022federated}
R.~S. Antunes, C.~Andr{\'e}~da Costa, A.~K{\"u}derle, I.~A. Yari, and
  B.~Eskofier, ``Federated learning for healthcare: Systematic review and
  architecture proposal,'' \emph{ACM Transactions on Intelligent Systems and
  Technology (TIST)}, vol.~13, no.~4, pp. 1--23, 2022.

\bibitem{liu2023online}
Q.~Liu, S.~Sun, M.~Liu, Y.~Wang, and B.~Gao, ``Online spatio-temporal
  correlation-based federated learning for traffic flow forecasting,''
  \emph{arXiv preprint arXiv:2302.08658}, 2023.

\bibitem{kanagavelu2021federated}
R.~Kanagavelu, Z.~Li, J.~Samsudin, S.~Hussain, F.~Yang, Y.~Yang, R.~S.~M. Goh,
  and M.~Cheah, ``Federated learning for advanced manufacturing based on
  industrial iot data analytics,'' \emph{Implementing Industry 4.0: The Model
  Factory as the Key Enabler for the Future of Manufacturing}, pp. 143--176,
  2021.

\bibitem{mcmahan2017communication}
B.~McMahan, E.~Moore, D.~Ramage, S.~Hampson, and B.~A. y~Arcas,
  ``Communication-efficient learning of deep networks from decentralized
  data,'' in \emph{Artificial intelligence and statistics}.\hskip 1em plus
  0.5em minus 0.4em\relax PMLR, 2017, pp. 1273--1282.

\bibitem{li2020federated}
T.~Li, A.~K. Sahu, M.~Zaheer, M.~Sanjabi, A.~Talwalkar, and V.~Smith,
  ``Federated optimization in heterogeneous networks,'' \emph{Proceedings of
  Machine Learning and Systems}, vol.~2, pp. 429--450, 2020.

\bibitem{kulkarni2020survey}
V.~Kulkarni, M.~Kulkarni, and A.~Pant, ``Survey of personalization techniques
  for federated learning,'' in \emph{2020 Fourth World Conference on Smart
  Trends in Systems, Security and Sustainability (WorldS4)}.\hskip 1em plus
  0.5em minus 0.4em\relax IEEE, 2020, pp. 794--797.

\bibitem{tan2022towards}
A.~Z. Tan, H.~Yu, L.~Cui, and Q.~Yang, ``Towards personalized federated
  learning,'' \emph{IEEE Transactions on Neural Networks and Learning Systems},
  2022.

\bibitem{t2020personalized}
C.~T~Dinh, N.~Tran, and J.~Nguyen, ``Personalized federated learning with
  moreau envelopes,'' \emph{Advances in Neural Information Processing Systems},
  vol.~33, pp. 21\,394--21\,405, 2020.

\bibitem{mills2021multi}
J.~Mills, J.~Hu, and G.~Min, ``Multi-task federated learning for personalised
  deep neural networks in edge computing,'' \emph{IEEE Transactions on Parallel
  and Distributed Systems}, vol.~33, no.~3, pp. 630--641, 2021.

\bibitem{jin2022personalized}
H.~Jin, D.~Bai, D.~Yao, Y.~Dai, L.~Gu, C.~Yu, and L.~Sun, ``Personalized edge
  intelligence via federated self-knowledge distillation,'' \emph{IEEE
  Transactions on Parallel and Distributed Systems}, vol.~34, no.~2, pp.
  567--580, 2022.

\bibitem{wu2022communication}
C.~Wu, F.~Wu, L.~Lyu, Y.~Huang, and X.~Xie, ``Communication-efficient federated
  learning via knowledge distillation,'' \emph{Nature communications}, vol.~13,
  no.~1, pp. 1--8, 2022.

\bibitem{sattler2021cfd}
F.~Sattler, A.~Marban, R.~Rischke, and W.~Samek, ``Cfd: Communication-efficient
  federated distillation via soft-label quantization and delta coding,''
  \emph{IEEE Transactions on Network Science and Engineering}, vol.~9, no.~4,
  pp. 2025--2038, 2021.

\bibitem{zhang2021parameterized}
J.~Zhang, S.~Guo, X.~Ma, H.~Wang, W.~Xu, and F.~Wu, ``Parameterized knowledge
  transfer for personalized federated learning,'' \emph{Advances in Neural
  Information Processing Systems}, vol.~34, pp. 10\,092--10\,104, 2021.

\bibitem{jee2023communication}
Y.~J. Cho, J.~Wang, T.~Chirvolu, and G.~Joshi, ``Communication-efficient and
  model-heterogeneous personalized federated learning via clustered knowledge
  transfer,'' \emph{IEEE Journal of Selected Topics in Signal Processing},
  vol.~17, no.~1, pp. 234--247, 2023.

\bibitem{wu2023fedict}
Z.~Wu, S.~Sun, Y.~Wang, M.~Liu, X.~Jiang, and B.~Gao, ``Fedict: Federated
  multi-task distillation for multi-access edge computing,'' \emph{arXiv
  preprint arXiv:2301.00389}, 2023.

\bibitem{zhu2021dataicml}
\BIBentryALTinterwordspacing
Z.~Zhu, J.~Hong, and J.~Zhou, ``Data-free knowledge distillation for
  heterogeneous federated learning,'' in \emph{Proceedings of the 38th
  International Conference on Machine Learning, {ICML} 2021, 18-24 July 2021,
  Virtual Event}, ser. Proceedings of Machine Learning Research, M.~Meila and
  T.~Zhang, Eds., vol. 139.\hskip 1em plus 0.5em minus 0.4em\relax {PMLR},
  2021, pp. 12\,878--12\,889. [Online]. Available:
  \url{http://proceedings.mlr.press/v139/zhu21b.html}
\BIBentrySTDinterwordspacing

\bibitem{jeong2018communication}
E.~Jeong, S.~Oh, H.~Kim, J.~Park, M.~Bennis, and S.-L. Kim,
  ``Communication-efficient on-device machine learning: Federated distillation
  and augmentation under non-iid private data,'' \emph{arXiv preprint
  arXiv:1811.11479}, 2018.

\bibitem{malkov2018efficient}
Y.~A. Malkov and D.~A. Yashunin, ``Efficient and robust approximate nearest
  neighbor search using hierarchical navigable small world graphs,'' \emph{IEEE
  transactions on pattern analysis and machine intelligence}, vol.~42, no.~4,
  pp. 824--836, 2018.

\bibitem{li2019fedmd}
D.~Li and J.~Wang, ``Fedmd: Heterogenous federated learning via model
  distillation,'' \emph{arXiv preprint arXiv:1910.03581}, 2019.

\bibitem{lecun1998gradient}
Y.~LeCun, L.~Bottou, Y.~Bengio, and P.~Haffner, ``Gradient-based learning
  applied to document recognition,'' \emph{Proceedings of the IEEE}, vol.~86,
  no.~11, pp. 2278--2324, 1998.

\bibitem{xiao2017fashion}
H.~Xiao, K.~Rasul, and R.~Vollgraf, ``Fashion-mnist: a novel image dataset for
  benchmarking machine learning algorithms,'' \emph{arXiv preprint
  arXiv:1708.07747}, 2017.

\bibitem{krizhevsky2009learning}
A.~Krizhevsky, G.~Hinton \emph{et~al.}, ``Learning multiple layers of features
  from tiny images,'' 2009.

\bibitem{darlow2018cinic}
L.~N. Darlow, E.~J. Crowley, A.~Antoniou, and A.~J. Storkey, ``Cinic-10 is not
  imagenet or cifar-10,'' \emph{arXiv preprint arXiv:1810.03505}, 2018.

\bibitem{yang2019federated}
Q.~Yang, Y.~Liu, T.~Chen, and Y.~Tong, ``Federated machine learning: Concept
  and applications,'' \emph{ACM Transactions on Intelligent Systems and
  Technology (TIST)}, vol.~10, no.~2, pp. 1--19, 2019.

\bibitem{reddi2021adaptive}
S.~Reddi, Z.~Charles, M.~Zaheer, Z.~Garrett, K.~Rush, J.~Kone{\v{c}}n{\`y},
  S.~Kumar, and H.~B. McMahan, ``Adaptive federated optimization,''
  \emph{International Conference on Learning Representations}, 2021.

\bibitem{guha2019one}
N.~Guha, A.~Talwalkar, and V.~Smith, ``One-shot federated learning,''
  \emph{arXiv preprint arXiv:1902.11175}, 2019.

\bibitem{zhang2022dense}
\BIBentryALTinterwordspacing
J.~Zhang, C.~Chen, B.~Li, L.~Lyu, S.~Wu, S.~Ding, C.~Shen, and C.~Wu,
  ``{DENSE}: Data-free one-shot federated learning,'' in \emph{Advances in
  Neural Information Processing Systems}, A.~H. Oh, A.~Agarwal, D.~Belgrave,
  and K.~Cho, Eds., 2022. [Online]. Available:
  \url{https://openreview.net/forum?id=QFQoxCFYEkA}
\BIBentrySTDinterwordspacing

\bibitem{he2020group}
C.~He, M.~Annavaram, and S.~Avestimehr, ``Group knowledge transfer: Federated
  learning of large cnns at the edge,'' \emph{arXiv preprint arXiv:2007.14513},
  2020.

\bibitem{wu2022exploring}
Z.~Wu, S.~Sun, M.~Liu, J.~Zhang, Y.~Wang, and Q.~Liu, ``Exploring the
  distributed knowledge congruence in proxy-data-free federated distillation,''
  \emph{arXiv preprint arXiv:2204.07028}, 2022.

\bibitem{itahara2021distill}
S.~Itahara, T.~Nishio, Y.~Koda, M.~Morikura, and K.~Yamamoto,
  ``Distillation-based semi-supervised federated learning for
  communication-efficient collaborative training with non-iid private data,''
  \emph{IEEE Transactions on Mobile Computing}, pp. 1--1, 2021.

\bibitem{liu2020client}
L.~Liu, J.~Zhang, S.~Song, and K.~B. Letaief, ``Client-edge-cloud hierarchical
  federated learning,'' in \emph{ICC 2020-2020 IEEE International Conference on
  Communications (ICC)}.\hskip 1em plus 0.5em minus 0.4em\relax IEEE, 2020, pp.
  1--6.

\bibitem{wang2022accelerating}
Z.~Wang, H.~Xu, J.~Liu, Y.~Xu, H.~Huang, and Y.~Zhao, ``Accelerating federated
  learning with cluster construction and hierarchical aggregation,'' \emph{IEEE
  Transactions on Mobile Computing}, 2022.

\bibitem{xu2022hierfedml}
Z.~Xu, D.~Zhao, W.~Liang, O.~F. Rana, P.~Zhou, M.~Li, W.~Xu, H.~Li, and Q.~Xia,
  ``Hierfedml: aggregator placement and ue assignment for hierarchical
  federated learning in mobile edge computing,'' \emph{IEEE Transactions on
  Parallel and Distributed Systems}, vol.~34, no.~1, pp. 328--345, 2022.

\bibitem{tan2020federated}
B.~Tan, B.~Liu, V.~Zheng, and Q.~Yang, ``A federated recommender system for
  online services,'' in \emph{Fourteenth ACM Conference on Recommender
  Systems}, 2020, pp. 579--581.

\bibitem{zhou2022sourcetarget}
\BIBentryALTinterwordspacing
X.~Zhou, Y.~Tian, and X.~Wang, ``Source-target unified knowledge distillation
  for memory-efficient federated domain adaptation on edge devices,'' 2022.
  [Online]. Available: \url{https://openreview.net/forum?id=8rCMq0yJMG}
\BIBentrySTDinterwordspacing

\bibitem{jiang2020customized}
H.~Jiang, M.~Liu, B.~Yang, Q.~Liu, J.~Li, and X.~Guo, ``Customized federated
  learning for accelerated edge computing with heterogeneous task targets,''
  \emph{Computer Networks}, vol. 183, p. 107569, 2020.

\bibitem{wu2023survey}
Z.~Wu, S.~Sun, Y.~Wang, M.~Liu, X.~Jiang, and R.~Li, ``Survey of knowledge
  distillation in federated edge learning,'' \emph{arXiv preprint
  arXiv:2301.05849}, 2023.

\bibitem{zhang2022fedzkt}
L.~Zhang, D.~Wu, and X.~Yuan, ``Fedzkt: Zero-shot knowledge transfer towards
  resource-constrained federated learning with heterogeneous on-device
  models,'' in \emph{2022 IEEE 42nd International Conference on Distributed
  Computing Systems (ICDCS)}.\hskip 1em plus 0.5em minus 0.4em\relax IEEE,
  2022, pp. 928--938.

\bibitem{yu2022resource}
S.~Yu, W.~Qian, and A.~Jannesari, ``Resource-aware federated learning using
  knowledge extraction and multi-model fusion,'' \emph{arXiv preprint
  arXiv:2208.07978}, 2022.

\bibitem{arivazhagan2019federated}
M.~G. Arivazhagan, V.~Aggarwal, A.~K. Singh, and S.~Choudhary, ``Federated
  learning with personalization layers,'' \emph{arXiv preprint
  arXiv:1912.00818}, 2019.

\bibitem{liu2022communication}
L.~Liu, J.~Zhang, S.~Song, and K.~B. Letaief, ``Communication-efficient
  federated distillation with active data sampling,'' in \emph{ICC 2022-IEEE
  International Conference on Communications}.\hskip 1em plus 0.5em minus
  0.4em\relax IEEE, 2022, pp. 201--206.

\end{thebibliography}


\begin{thebibliography}{10}
\providecommand{\url}[1]{#1}
\csname url@samestyle\endcsname
\providecommand{\newblock}{\relax}
\providecommand{\bibinfo}[2]{#2}
\providecommand{\BIBentrySTDinterwordspacing}{\spaceskip=0pt\relax}
\providecommand{\BIBentryALTinterwordstretchfactor}{4}
\providecommand{\BIBentryALTinterwordspacing}{\spaceskip=\fontdimen2\font plus
\BIBentryALTinterwordstretchfactor\fontdimen3\font minus \fontdimen4\font\relax}
\providecommand{\BIBforeignlanguage}[2]{{%
\expandafter\ifx\csname l@#1\endcsname\relax
\typeout{** WARNING: IEEEtran.bst: No hyphenation pattern has been}%
\typeout{** loaded for the language `#1'. Using the pattern for}%
\typeout{** the default language instead.}%
\else
\language=\csname l@#1\endcsname
\fi
#2}}
\providecommand{\BIBdecl}{\relax}
\BIBdecl

\bibitem{iot-predict}
\BIBentryALTinterwordspacing
statista, ``Number of internet of things (iot) connections worldwide from 2022 to 2023, with forecasts from 2024 to 2033.'' [Online]. Available: \url{https://www.statista.com/statistics/1183457/iot-connected-devices-worldwide/}
\BIBentrySTDinterwordspacing

\bibitem{baccour2022pervasive}
E.~Baccour, N.~Mhaisen, A.~A. Abdellatif, A.~Erbad, A.~Mohamed, M.~Hamdi, and M.~Guizani, ``Pervasive ai for iot applications: A survey on resource-efficient distributed artificial intelligence,'' \emph{IEEE Communications Surveys \& Tutorials}, vol.~24, no.~4, pp. 2366--2418, 2022.

\bibitem{yang2019federated}
Q.~Yang, Y.~Liu, T.~Chen, and Y.~Tong, ``Federated machine learning: Concept and applications,'' \emph{ACM Transactions on Intelligent Systems and Technology (TIST)}, vol.~10, no.~2, pp. 1--19, 2019.

\bibitem{nguyen2021federated}
D.~C. Nguyen, M.~Ding, P.~N. Pathirana, A.~Seneviratne, J.~Li, and H.~V. Poor, ``Federated learning for internet of things: A comprehensive survey,'' \emph{IEEE Communications Surveys \& Tutorials}, vol.~23, no.~3, pp. 1622--1658, 2021.

\bibitem{mcmahan2017communication}
B.~McMahan, E.~Moore, D.~Ramage, S.~Hampson, and B.~A. y~Arcas, ``Communication-efficient learning of deep networks from decentralized data,'' in \emph{Artificial intelligence and statistics}.\hskip 1em plus 0.5em minus 0.4em\relax PMLR, 2017, pp. 1273--1282.

\bibitem{li2020federated}
T.~Li, A.~K. Sahu, M.~Zaheer, M.~Sanjabi, A.~Talwalkar, and V.~Smith, ``Federated optimization in heterogeneous networks,'' \emph{Proceedings of Machine Learning and Systems}, vol.~2, pp. 429--450, 2020.

\bibitem{karimireddy2020scaffold}
S.~P. Karimireddy, S.~Kale, M.~Mohri, S.~Reddi, S.~Stich, and A.~T. Suresh, ``Scaffold: Stochastic controlled averaging for federated learning,'' in \emph{International conference on machine learning}.\hskip 1em plus 0.5em minus 0.4em\relax PMLR, 2020, pp. 5132--5143.

\bibitem{duan2022distributed}
S.~Duan, D.~Wang, J.~Ren, F.~Lyu, Y.~Zhang, H.~Wu, and X.~Shen, ``Distributed artificial intelligence empowered by end-edge-cloud computing: A survey,'' \emph{IEEE Communications Surveys \& Tutorials}, vol.~25, no.~1, pp. 591--624, 2022.

\bibitem{wang2024end}
Y.~Wang, C.~Yang, S.~Lan, L.~Zhu, and Y.~Zhang, ``End-edge-cloud collaborative computing for deep learning: A comprehensive survey,'' \emph{IEEE Communications Surveys \& Tutorials}, 2024.

\bibitem{bao2022federated}
G.~Bao and P.~Guo, ``Federated learning in cloud-edge collaborative architecture: key technologies, applications and challenges,'' \emph{Journal of Cloud Computing}, vol.~11, no.~1, p.~94, 2022.

\bibitem{liu2020client}
L.~Liu, J.~Zhang, S.~Song, and K.~B. Letaief, ``Client-edge-cloud hierarchical federated learning,'' in \emph{ICC 2020-2020 IEEE International Conference on Communications (ICC)}.\hskip 1em plus 0.5em minus 0.4em\relax IEEE, 2020, pp. 1--6.

\bibitem{tak2020federated}
A.~Tak and S.~Cherkaoui, ``Federated edge learning: Design issues and challenges,'' \emph{IEEE Network}, vol.~35, no.~2, pp. 252--258, 2020.

\bibitem{wang2023accelerating}
Z.~Wang, H.~Xu, J.~Liu, Y.~Xu, H.~Huang, and Y.~Zhao, ``Accelerating federated learning with cluster construction and hierarchical aggregation,'' \emph{IEEE Transactions on Mobile Computing}, vol.~22, no.~7, pp. 3805--3822, 2023.

\bibitem{yang2023hierarchical}
Z.~Yang, S.~Fu, W.~Bao, D.~Yuan, and A.~Y. Zomaya, ``Hierarchical federated learning with momentum acceleration in multi-tier networks,'' \emph{IEEE Transactions on Parallel and Distributed Systems}, 2023.

\bibitem{liu2022hierarchical}
L.~Liu, J.~Zhang, S.~Song, and K.~B. Letaief, ``Hierarchical federated learning with quantization: Convergence analysis and system design,'' \emph{IEEE Transactions on Wireless Communications}, vol.~22, no.~1, pp. 2--18, 2023.

\bibitem{wang2019dynamic}
S.~Wang, R.~Urgaonkar, M.~Zafer, T.~He, K.~Chan, and K.~K. Leung, ``Dynamic service migration in mobile edge computing based on markov decision process,'' \emph{IEEE/ACM Transactions on Networking}, vol.~27, no.~3, pp. 1272--1288, 2019.

\bibitem{jsac-dynamic}
T.~Ouyang, Z.~Zhou, and X.~Chen, ``Follow me at the edge: Mobility-aware dynamic service placement for mobile edge computing,'' \emph{IEEE Journal on Selected Areas in Communications}, vol.~36, no.~10, pp. 2333--2345, 2018.

\bibitem{deng2023hierarchical}
Y.~Deng, J.~Ren, C.~Tang, F.~Lyu, Y.~Liu, and Y.~Zhang, ``A hierarchical knowledge transfer framework for heterogeneous federated learning,'' in \emph{IEEE INFOCOM 2023-IEEE Conference on Computer Communications}.\hskip 1em plus 0.5em minus 0.4em\relax IEEE, 2023, pp. 1--10.

\bibitem{makd}
\BIBentryALTinterwordspacing
A.~Afonin and S.~P. Karimireddy, ``Towards model agnostic federated learning using knowledge distillation,'' in \emph{International Conference on Learning Representations}, 2022. [Online]. Available: \url{https://openreview.net/forum?id=lQI_mZjvBxj}
\BIBentrySTDinterwordspacing

\bibitem{cheng2021fedgems}
S.~Cheng, J.~Wu, Y.~Xiao, and Y.~Liu, ``Fedgems: Federated learning of larger server models via selective knowledge fusion,'' \emph{arXiv preprint arXiv:2110.11027}, 2021.

\bibitem{cho2022heterogeneous}
\BIBentryALTinterwordspacing
Y.~J. Cho, A.~Manoel, G.~Joshi, R.~Sim, and D.~Dimitriadis, ``Heterogeneous ensemble knowledge transfer for training large models in federated learning,'' in \emph{Proceedings of the Thirty-First International Joint Conference on Artificial Intelligence, {IJCAI-22}}, L.~D. Raedt, Ed.\hskip 1em plus 0.5em minus 0.4em\relax International Joint Conferences on Artificial Intelligence Organization, 7 2022, pp. 2881--2887, main Track. [Online]. Available: \url{https://doi.org/10.24963/ijcai.2022/399}
\BIBentrySTDinterwordspacing

\bibitem{alam2022fedrolex}
S.~Alam, L.~Liu, M.~Yan, and M.~Zhang, ``Fedrolex: Model-heterogeneous federated learning with rolling sub-model extraction,'' \emph{Advances in neural information processing systems}, vol.~35, pp. 29\,677--29\,690, 2022.

\bibitem{he2020group}
C.~He, M.~Annavaram, and S.~Avestimehr, ``Group knowledge transfer: Federated learning of large cnns at the edge,'' \emph{Advances in Neural Information Processing Systems}, vol.~33, pp. 14\,068--14\,080, 2020.

\bibitem{wu2022exploring}
Z.~Wu, S.~Sun, Y.~Wang, M.~Liu, Q.~Pan, J.~Zhang, Z.~Li, and Q.~Liu, ``Exploring the distributed knowledge congruence in proxy-data-free federated distillation,'' \emph{ACM Transactions on Intelligent Systems and Technology}, vol.~15, no.~2, pp. 1--34, 2024.

\bibitem{nguyen2023self}
M.~N.~H. Nguyen, S.~R. Pandey, T.~N. Dang, E.-N. Huh, N.~H. Tran, W.~Saad, and C.~S. Hong, ``Self-organizing democratized learning: Toward large-scale distributed learning systems,'' \emph{IEEE Transactions on Neural Networks and Learning Systems}, vol.~34, no.~12, pp. 10\,698--10\,710, 2023.

\bibitem{wu2024agglomerative}
Z.~Wu, S.~Sun, Y.~Wang, M.~Liu, B.~Gao, Q.~Pan, T.~He, and X.~Jiang, ``Agglomerative federated learning: Empowering larger model training via end-edge-cloud collaboration,'' in \emph{IEEE INFOCOM 2024-IEEE Conference on Computer Communications}.\hskip 1em plus 0.5em minus 0.4em\relax IEEE, 2024, pp. 131--140.

\bibitem{wang2022accelerating}
Z.~Wang, H.~Xu, J.~Liu, Y.~Xu, H.~Huang, and Y.~Zhao, ``Accelerating federated learning with cluster construction and hierarchical aggregation,'' \emph{IEEE Transactions on Mobile Computing}, vol.~22, no.~7, pp. 3805--3822, 2023.

\bibitem{anil2018large}
R.~Anil, G.~Pereyra, A.~Passos, R.~Ormandi, G.~E. Dahl, and G.~E. Hinton, ``Large scale distributed neural network training through online distillation,'' \emph{arXiv preprint arXiv:1804.03235}, 2018.

\bibitem{wang2021knowledge}
L.~Wang and K.-J. Yoon, ``Knowledge distillation and student-teacher learning for visual intelligence: A review and new outlooks,'' \emph{IEEE Transactions on Pattern Analysis and Machine Intelligence}, 2021.

\bibitem{deng2009imagenet}
J.~Deng, W.~Dong, R.~Socher, L.-J. Li, K.~Li, and L.~Fei-Fei, ``Imagenet: A large-scale hierarchical image database,'' in \emph{2009 IEEE conference on computer vision and pattern recognition}.\hskip 1em plus 0.5em minus 0.4em\relax Ieee, 2009, pp. 248--255.

\bibitem{nguyen2023feddct}
Q.~Nguyen, H.~H. Pham, K.-S. Wong, P.~Le~Nguyen, T.~T. Nguyen, and M.~N. Do, ``Feddct: Federated learning of large convolutional neural networks on resource constrained devices using divide and collaborative training,'' \emph{IEEE Transactions on Network and Service Management}, 2023.

\bibitem{he2020fedml}
C.~He, S.~Li, J.~So, X.~Zeng, M.~Zhang, H.~Wang, X.~Wang, P.~Vepakomma, A.~Singh, H.~Qiu \emph{et~al.}, ``Fedml: A research library and benchmark for federated machine learning,'' \emph{arXiv preprint arXiv:2007.13518}, 2020.

\bibitem{netzer2011reading}
Y.~Netzer, T.~Wang, A.~Coates, A.~Bissacco, B.~Wu, A.~Y. Ng \emph{et~al.}, ``Reading digits in natural images with unsupervised feature learning,'' in \emph{NIPS workshop on deep learning and unsupervised feature learning}, vol. 2011, no.~2.\hskip 1em plus 0.5em minus 0.4em\relax Granada, 2011, p.~4.

\bibitem{cifar10}
A.~Krizhevsky, G.~Hinton \emph{et~al.}, ``Learning multiple layers of features from tiny images,'' 2009.

\bibitem{cinic10}
L.~N. Darlow, E.~J. Crowley, A.~Antoniou, and A.~J. Storkey, ``Cinic-10 is not imagenet or cifar-10,'' \emph{arXiv preprint arXiv:1810.03505}, 2018.

\bibitem{he2016deep}
K.~He, X.~Zhang, S.~Ren, and J.~Sun, ``Deep residual learning for image recognition,'' in \emph{Proceedings of the IEEE conference on computer vision and pattern recognition}, 2016, pp. 770--778.

\bibitem{li2022federated}
Q.~Li, Y.~Diao, Q.~Chen, and B.~He, ``Federated learning on non-iid data silos: An experimental study,'' in \emph{2022 IEEE 38th international conference on data engineering (ICDE)}.\hskip 1em plus 0.5em minus 0.4em\relax IEEE, 2022, pp. 965--978.

\bibitem{demlearn-code}
\BIBentryALTinterwordspacing
M.~N. Nguyen, S.~R. Pandey, T.~N. Dang, E.-N. Huh, N.~H. Tran, W.~Saad, and C.~S. Hong. [Online]. Available: \url{https://github.com/nhatminh/Dem-AI}
\BIBentrySTDinterwordspacing

\bibitem{hierqsgd-code}
\BIBentryALTinterwordspacing
L.~Liu, J.~Zhang, S.~Song, and K.~B. Letaief. [Online]. Available: \url{https://github.com/LuminLiu/Hier_QSGD}
\BIBentrySTDinterwordspacing

\bibitem{duan2023combining}
Q.~Duan, J.~Huang, S.~Hu, R.~Deng, Z.~Lu, and S.~Yu, ``Combining federated learning and edge computing toward ubiquitous intelligence in 6g network: Challenges, recent advances, and future directions,'' \emph{IEEE Communications Surveys \& Tutorials}, 2023.

\bibitem{liu2023hierarchical}
L.~Liu, J.~Zhang, S.~Song, and K.~B. Letaief, ``Hierarchical federated learning with quantization: Convergence analysis and system design,'' \emph{IEEE Transactions on Wireless Communications}, vol.~22, no.~1, pp. 2--18, 2023.

\bibitem{guo2023privacy}
Y.~Guo, F.~Liu, T.~Zhou, Z.~Cai, and N.~Xiao, ``Privacy vs. efficiency: Achieving both through adaptive hierarchical federated learning,'' \emph{IEEE Transactions on Parallel and Distributed Systems}, vol.~34, no.~4, pp. 1331--1342, 2023.

\bibitem{wu2023survey}
Z.~Wu, S.~Sun, Y.~Wang, M.~Liu, X.~Jiang, and R.~Li, ``Survey of knowledge distillation in federated edge learning,'' \emph{arXiv preprint arXiv:2301.05849}, 2023.

\bibitem{wu2023fedict}
Z.~Wu, S.~Sun, Y.~Wang, M.~Liu, Q.~Pan, X.~Jiang, and B.~Gao, ``Fedict: Federated multi-task distillation for multi-access edge computing,'' \emph{IEEE Transactions on Parallel and Distributed Systems}, 2023.

\bibitem{jin2023personalized}
H.~Jin, D.~Bai, D.~Yao, Y.~Dai, L.~Gu, C.~Yu, and L.~Sun, ``Personalized edge intelligence via federated self-knowledge distillation,'' \emph{IEEE Transactions on Parallel and Distributed Systems}, vol.~34, no.~2, pp. 567--580, 2023.

\bibitem{yao2024fedgkd}
D.~Yao, W.~Pan, Y.~Dai, Y.~Wan, X.~Ding, C.~Yu, H.~Jin, Z.~Xu, and L.~Sun, ``Fedgkd: Toward heterogeneous federated learning via global knowledge distillation,'' \emph{IEEE Transactions on Computers}, vol.~73, no.~1, pp. 3--17, 2024.

\bibitem{liu2024adaptive}
J.~Liu, Q.~Zeng, H.~Xu, Y.~Xu, Z.~Wang, and H.~Huang, ``Adaptive block-wise regularization and knowledge distillation for enhancing federated learning,'' \emph{IEEE/ACM Transactions on Networking}, vol.~32, no.~1, pp. 791--805, 2024.

\bibitem{wu2024fedcache}
Z.~Wu, S.~Sun, Y.~Wang, M.~Liu, K.~Xu, W.~Wang, X.~Jiang, B.~Gao, and J.~Lu, ``Fedcache: A knowledge cache-driven federated learning architecture for personalized edge intelligence,'' \emph{IEEE Transactions on Mobile Computing}, 2024.

\bibitem{pan2024fedcache}
Q.~Pan, S.~Sun, Z.~Wu, Y.~Wang, M.~Liu, and B.~Gao, ``Fedcache 2.0: Exploiting the potential of distilled data in knowledge cache-driven federated learning,'' \emph{arXiv preprint arXiv:2405.13378}, 2024.

\end{thebibliography}

\vspace{-33pt}
 \begin{IEEEbiography}[{\includegraphics[width=1in,height=1.25in,clip,keepaspectratio]{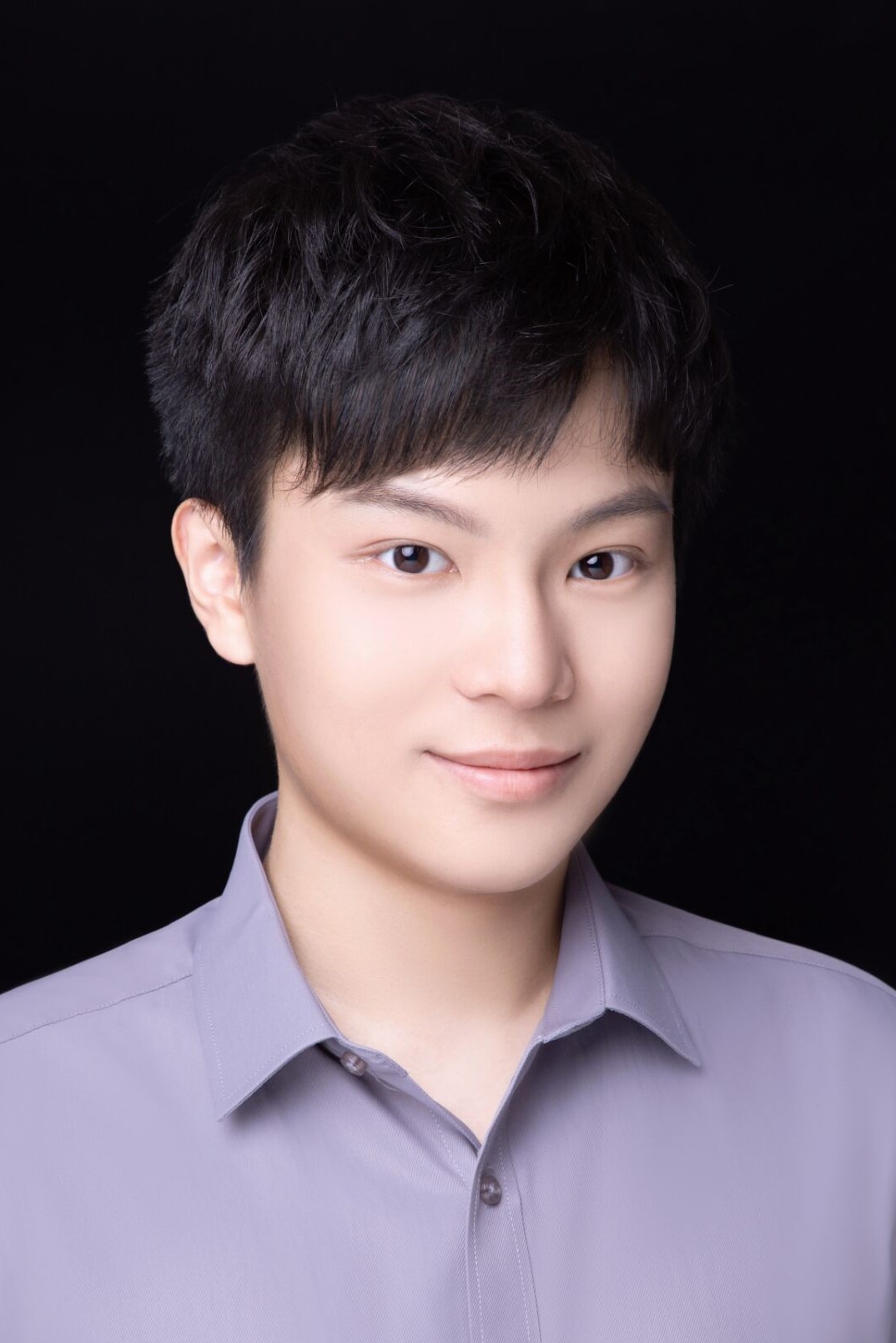}}]{Zhiyuan Wu} (Member, IEEE) is currently a research assistant with the Institute of Computing Technology, Chinese Academy of Sciences (ICT, CAS). He has contributed several technical papers to top-tier conferences and journals as the first author in the fields of computer architecture, computer networks, and intelligent systems, including IEEE Transactions on Parallel and Distributed Systems (TPDS), IEEE Transactions on Mobile Computing (TMC), IEEE International Conference on Computer Communications (INFOCOM), and ACM Transactions on Intelligent Systems and Technology (TIST). He is a member of IEEE, ACM, the China Computer Federation (CCF), and is granted the President Special Prize of ICT, CAS. His research interests include federated learning, mobile edge computing, and distributed systems.
\end{IEEEbiography}
 
 \begin{IEEEbiography}[{\includegraphics[width=1in,height=1.25in,clip,keepaspectratio]{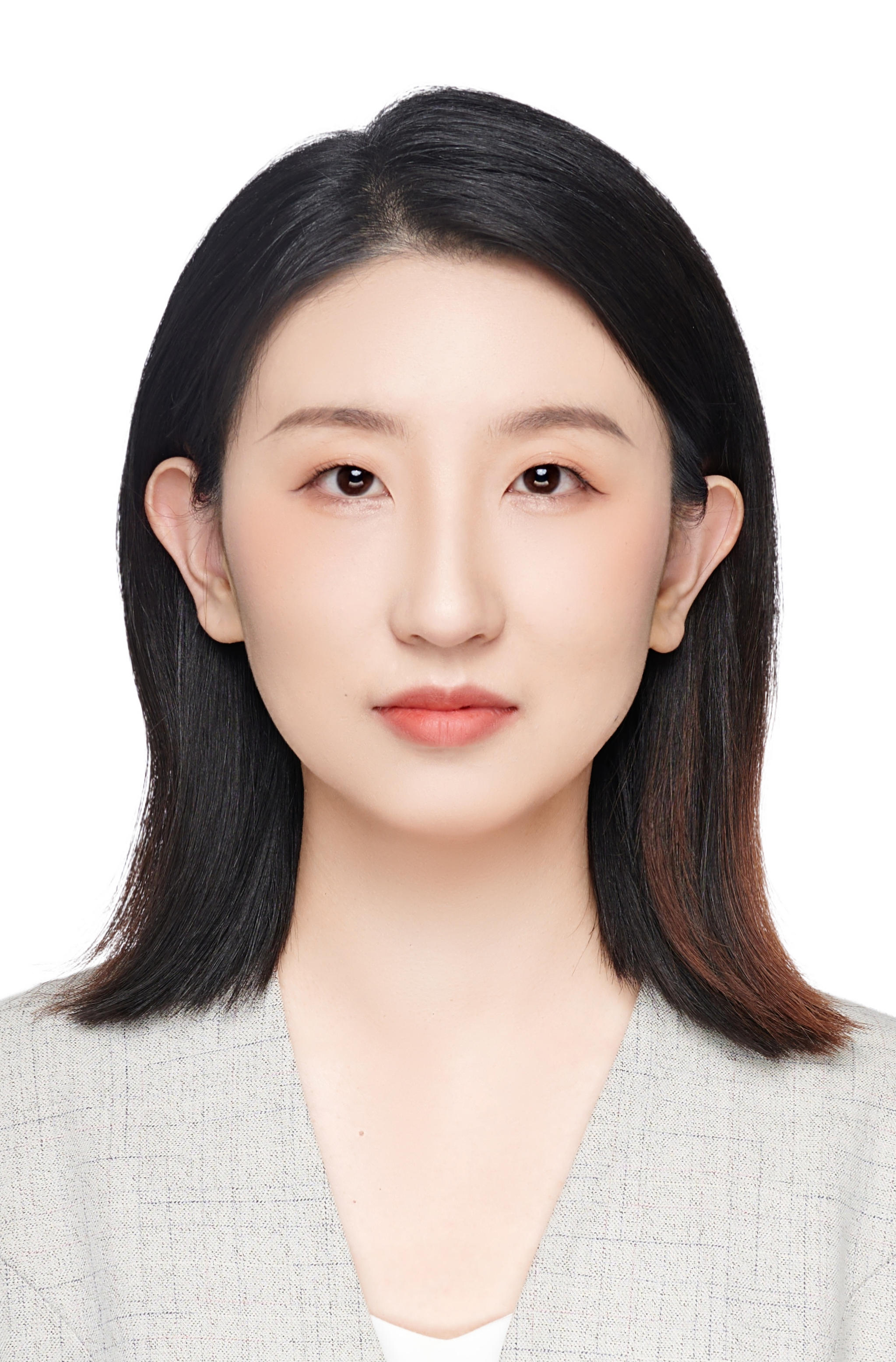}}]{Sheng Sun} is currently an associate professor at the Institute of Computing Technology, Chinese Academy of Sciences. She received her bachelor's degree from Beihang University, and her Ph.D. from the Institute of Computing Technology, Chinese Academy of Sciences. Dr. Sun has led or executed 5 major funded research projects and published over 20 technical papers in journals and conferences related to computer network and distributed systems, including IEEE Transactions on Parallel and Distributed Systems (TPDS), IEEE Transactions on Mobile Computing (TMC), and IEEE International Conference on Computer Communications (INFOCOM). Her research interests include federated learning, edge intelligence, and privacy computing.
\end{IEEEbiography}

\begin{IEEEbiography}[{\includegraphics[width=1in,height=1.25in,clip,keepaspectratio]{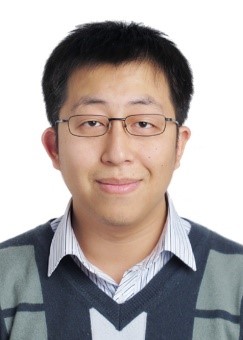}}]{Yuwei Wang} (Member, IEEE) received his Ph.D. degree in computer science from the University of Chinese Academy of Sciences, Beijing, China. He is currently an associate professor at the Institute of Computing Technology, Chinese Academy of Sciences. He has been responsible for setting over 30 international and national standards, and also holds various positions in both international and national industrial standards development organizations (SDOs) as well as local research institutions, including the associate rapporteur at the ITU-T SG21 Q5, and the deputy director of China Communications Standards Association (CCSA) TC1 WG1. His current research interests include federated learning, mobile edge computing, and next-generation network architecture.
\end{IEEEbiography}

\begin{IEEEbiography}[{\includegraphics[width=1in,height=1.25in,clip,keepaspectratio]{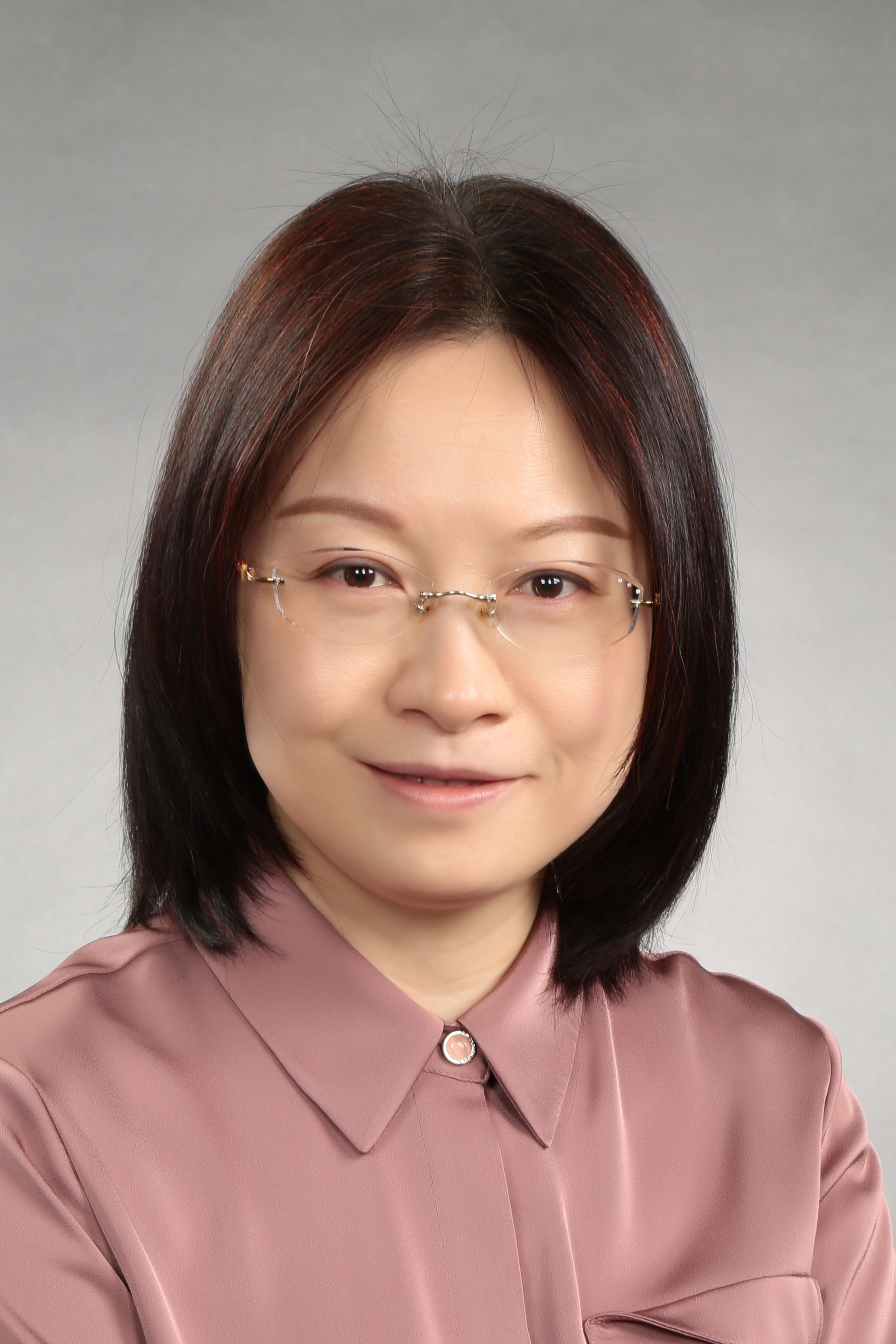}}]{Min Liu} (Senior Member, IEEE) received her Ph.D degree in computer science from the Graduate University of the Chinese Academy of Sciences, China. Before that, she received her B.S. and M.S. degrees in computer science from Xi’an Jiaotong University, China. She is currently a professor at the Institute of Computing Technology, Chinese Academy of Sciences, and also holds a position at the Zhongguancun Laboratory. Her current research interests include mobile computing and edge intelligence.
\end{IEEEbiography}

\begin{IEEEbiography}[{\includegraphics[width=1in,height=1.25in,clip,keepaspectratio]{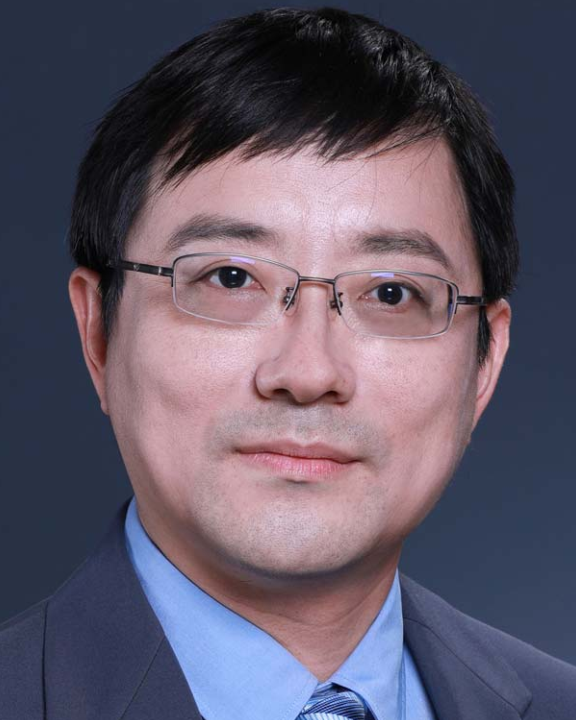}}]{Ke Xu}
	(Fellow, IEEE) received the Ph.D. degree from the Department of Computer Science and Technology, Tsinghua University, Beijing, China. He serves as a Full Professor at Tsinghua University. He has published more than 200 technical papers and holds 11 U.S. patents in the research areas of next-generation internet, blockchain systems, the Internet of Things, and network security. He is a member of ACM. He served as the Steering Committee Chair for IEEE/ACM IWQoS.
He has guest-edited several special issues in IEEE and Springer journals. He is the Editor of IEEE Internet of Things Journal.
\end{IEEEbiography}

\begin{IEEEbiography}[{\includegraphics[width=1in,height=1.25in,clip,keepaspectratio]{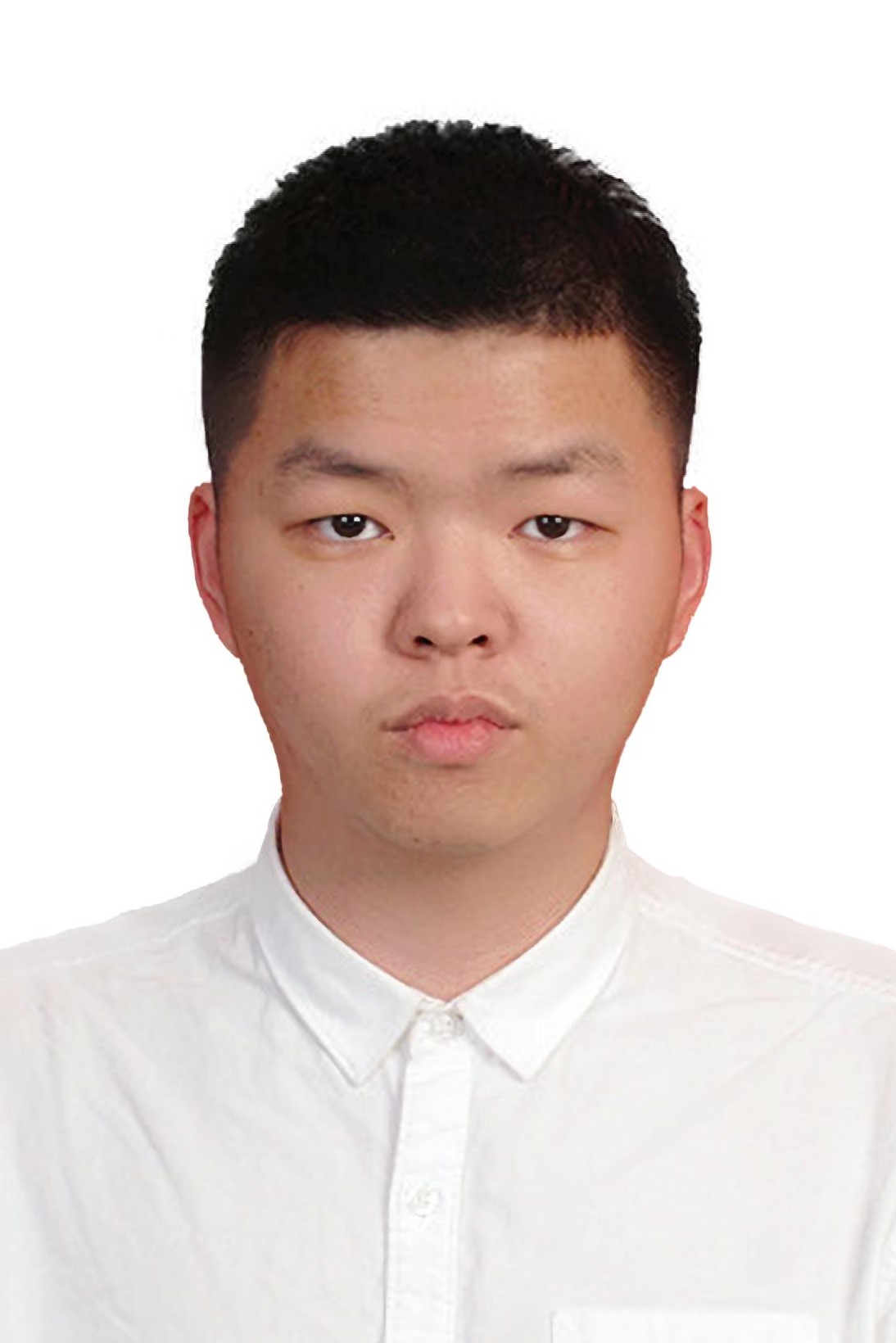}}]{Quyang Pan}
	is currently a research assistant with the Institute of Computing Technology, Chinese Academy of Sciences. He has published several technical papers at top-tier conferences and journals in the fields of computer architecture, computer networks, and intelligent systems, including IEEE Transactions on Parallel and Distributed Systems (TPDS), IEEE Transactions on Mobile Computing (TMC), IEEE International Conference on Computer Communications (INFOCOM), and ACM Transactions on Intelligent Systems and Technology (TIST). He is also an outstanding competitive programmer who has won several gold medals in international and national contests such as ACM International Collegiate Programming Contest (ICPC), CCF Collegiate Computer Systems and Programming Contest (CCSP), etc. His research interests include federated learning and edge computing.
\end{IEEEbiography}



\begin{IEEEbiography}[{\includegraphics[width=1in,height=1.25in,clip,keepaspectratio]{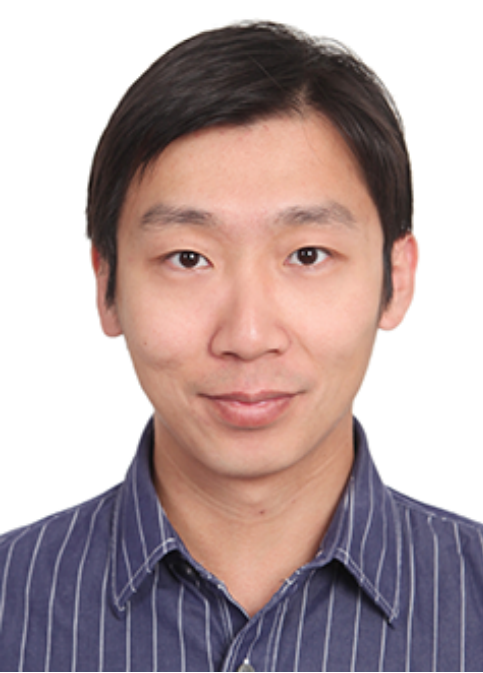}}]{Bo Gao} (Member, IEEE) received his M.S. degree in electrical engineering from the School of Electronic Information and Electrical Engineering at Shanghai Jiaotong University, Shanghai, China in 2009, and his Ph.D. degree in computer engineering from the Bradley Department of Electrical and Computer Engineering at Virginia Tech, Blacksburg, USA in 2014. He was an Assistant Professor with the Institute of Computing Technology at Chinese Academy of Sciences, Beijing, China from 2014 to 2017. He was a Visiting Researcher with the School of Computing and Communications at Lancaster University, Lancaster, UK from 2018 to 2019. He is currently an Associate Professor with the School of Computer and Information Technology at Beijing Jiaotong University, Beijing, China. He has directed a number of research projects sponsored by the National Natural Science Foundation of China (NSFC) or other funding agencies. He is a member of IEEE, ACM, and China Computer Federation (CCF). His research interests include wireless networking, mobile/edge computing, multiagent systems, and machine learning.
\end{IEEEbiography}

\begin{IEEEbiography}[{\includegraphics[width=1in,height=1.25in,clip,keepaspectratio]{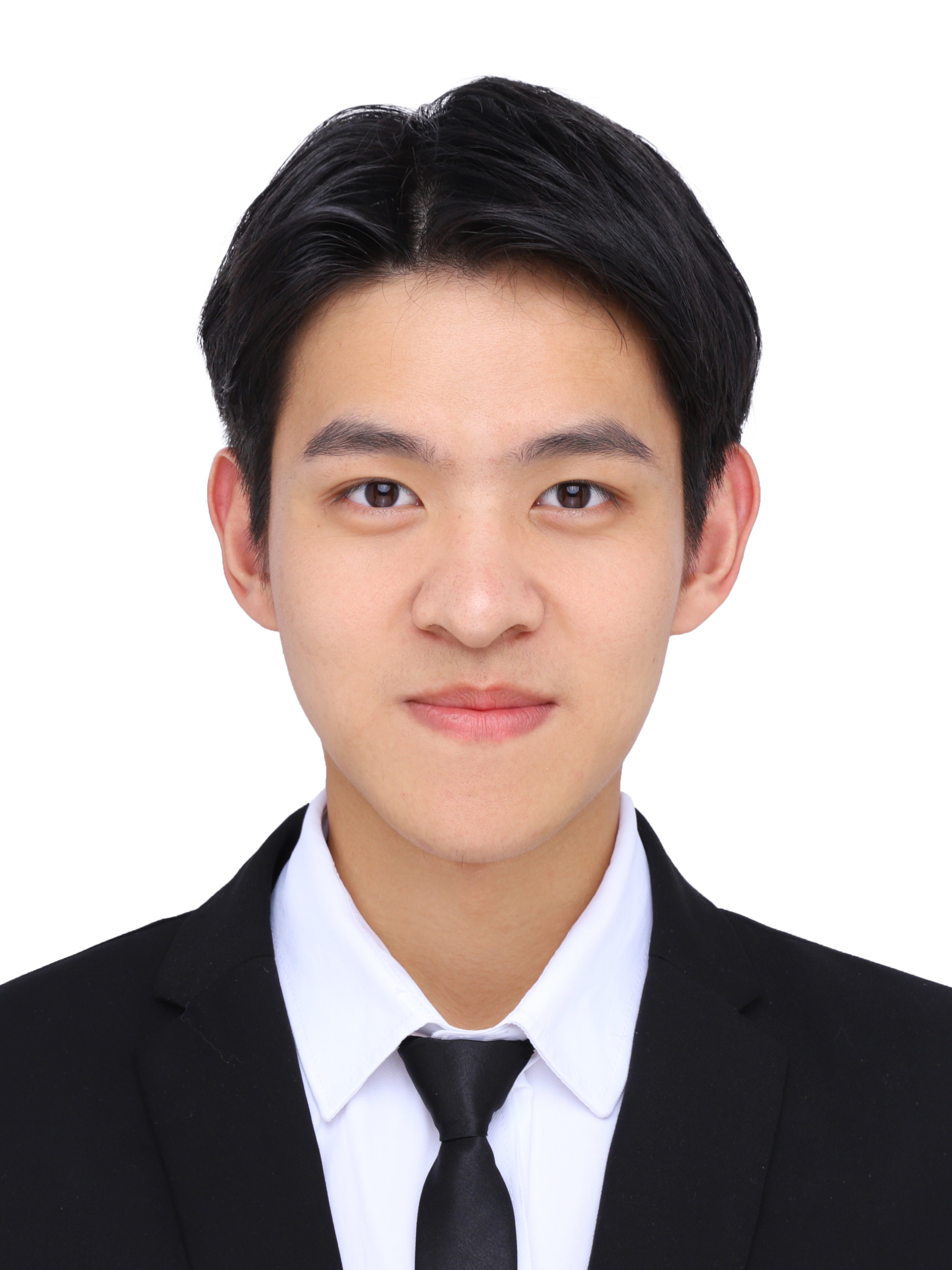}}]{Tian Wen} is currently a research assistant with the Institute of Computing Technology, Chinese Academy of Sciences. His research interests include federated learning, edge computing, and information security.
\end{IEEEbiography}



\vfill
\newpage
\appendices
\onecolumn
\end{document}